\documentclass[a4paper,12pt]{article}
\usepackage{amsthm}
\usepackage{listings}
\usepackage{verbatim}
\usepackage{amsmath,amssymb}
\usepackage{amsfonts}
\usepackage{graphicx}
\usepackage{pdfpages}
\usepackage{caption}
\usepackage{subcaption}
\usepackage[english]{babel}
\usepackage{hyperref}
\usepackage{float}
\usepackage[T1]{fontenc}
\usepackage[utf8]{inputenc}
\usepackage{authblk}

\usepackage{perpage} 
\usepackage{chngcntr}
\theoremstyle{definition}
\usepackage[titletoc,title]{appendix}
\usepackage{tocloft}
\usepackage[all]{xy}
\usepackage{sectsty}
\usepackage[top=25mm, bottom=25mm, left=15mm, right=15mm]{geometry}

\begin{document}

\title{Constructing a partially transparent computational boundary for UPPE using leaky modes}
\author[*]{David Juhasz}
\author[*]{Per Kristen Jakobsen}
\affil[*]{Department of Mathematics and Statistics, the Arctic University of Norway, 9019 Troms\o, Norway}
\date{\today}

\maketitle
\begin{abstract}
In this paper we introduce a method for creating a transparent computational boundary for
the simulation of unidirectional propagation of optical beams and pulses using leaky modes.
The key element of the method is the introduction of an artificial-index material outside a chosen
computational domain and utilization of the quasi-normal modes associated with such artificial structure.
The method is tested on the free space propagation of TE electromagnetic waves. By choosing the material
to have appropriate optical properties one can greatly reduce the reflection at the computational boundary.
In contrast to the well-known approach based on a perfectly matched layer, our method is
especially well suited for spectral propagators.
\end{abstract}

\setcounter{figure}{0}
\setcounter{page}{1}
\pagenumbering{arabic}
\section[Introduction]{Introduction}

Treatment of domain boundaries in numerical simulations, especially in the solution of partial differential
equations, presents a long-standing problem. While powerful methods have been developed for certain
situations, they often introduce significant additional complexity and computational overhead. The perfectly
matched layer (PML)~\cite{Berenger94} approach stands as a prime example of methods that work extremely well in situations
where a transparent boundary is meant to mimic a connection of the given ``computational box'' to an
infinite outside space. Indeed, PML-based methods are routinely employed for wave-propagation simulation,
for example in finite-difference Maxwell solvers~\cite{CoordStretch} and in beam-propagation simulation~\cite{BPMPML}.

Nevertheless, there are applications for which good boundary treatments are still lacking. For example,
in extreme nonlinear optics, characterized by high intensity, few cycle pulses, which, through their
interaction with material degrees of freedom, display very broad and complex spatio-temporal spectra,
{\em spectral beam and pulse propagators}~\cite{Guide} are the preferred methods of choice. Unfortunately they do
not mesh well with the boundary treatments developed for the finite-difference solvers such as PML.

While spectral propagators applied to pulses and/or beams shine in many situations that are next to
impossible to handle with finite-difference approaches, the boundary treatment can be a significant
problem. For example, long-distance propagation of highly nonlinear optical pulses~\cite{Schuh:17} is often connected
with light-matter interactions that send significant energy propagating toward the boundaries of
computational domains where it must be ``absorbed'' as if propagating into infinite space. In connection
to spectral-based numerical simulation of beam and pulse propagation, this is a difficult problem that
we aim to address in this work.

The method we put forward can be understood as an extension of an approximation that is
sometimes used to simulate beam and pulse propagation in leaky waveguides~\cite{Chen:08,Chen:11} such as hollow-core
fibers or capillaries~\cite{Popmintchev1287}. In such a context, the
propagating modes are approximated~\cite{Arnold:09,Arnold} by real parts of the true leaky modes for the given waveguide~\cite{Marcatili},
while their propagation constants are redefined by inclusion of the imaginary parts that reflect the
propagation loss of a leaky mode. Such an approach can be interpreted  as a first-order
perturbation theory where eigenvalues are corrected while the wave functions are kept unchanged.
Needless to say, this only works when the physics dictates that the propagation is dominated
by a relatively small number of modes that have small propagation losses.

We propose to utilize the true leaky modes, without approximations, as the basis for both the
numerical representation of the optical field and for the realization of transparent boundary
conditions. We introduce an artificial structure outside of the given computational domain
in order to introduce an infinite set of quasi-normal modes, and construct an expansion of an arbitrary
beam profile. While we present the treatment for a fixed frequency, the generalization to
pulsed waveform is straightforward.

Leaky modes has had a long history in the field of electromagnetics. They were used already as  early as in 1884  by J. J. Thomson\cite{Thomson} in his study of decay phenomena in electromagnetics.  Since then, they have been of enduring interest  in electromagnetics, for resonator cavities~\cite{deLasson:14,Kristensen14},
optical waveguides~\cite{Franchimon:13},  photonic~\cite{Settimi:09} and plasmonic ~\cite{Ge:15,Fan16} structures, and are often used for  numerical
simulations, which is also what we propose to do in the current  paper. Leaky modes are decaying eigenstates and as such have played an important role in quantum theory from its very inception until today. In this setting they describe unstable states. Such states were first defined in terms of  the absence of incoming waves  by Siegert\cite{Siegert} for the nuclear scattering matrix.  Siegert's definition of unstable states  was taken up by Peierls\cite{Peierls},Couteur\cite{Couteur} and Humblet\cite{Humblet} and by them refined  into an important tool for nuclear scattering theory. The wave functions satisfying the Siegert outgoing-wave conditions are known as resonant states, and their properties has been of interest for many years
 \cite{Goto,Lind2,garcia1,siegert1,siegert2,siegert3}. 

As is evident from the previous paragraph, leaky modes and unstable states have  a long history and has been, and is, of great  utility \cite{tolstikhin,resonance_applications,madrid,moiseyev_review}
in various fields. However, the fact that they are decaying eigenstates means that the corresponding eigenvalue problems are not self adjoint. Consequently, the matter of projecting general field configurations  into sum of leaky modes or resonant states, and the question of completeness of the resulting expansions, are not backed up by any general theory,  like for the self adjoint case.
In fact the leaky modes and resonant states are invariably growing exponentially in space, and thus can not be placed  in some well known inner product spaces.  The lack of a general theory for non-self adjoint operators is challenging, and it means that questions of projection and completeness have to be handled in a case by case basis.  In this paper we will introduce a projection method for leaky modes based on a naturally occurring  complex non-Hermitian inner product, but will not present a convergence proof for our leaky mode expansions.

The paper is organized as follows. In section two we introduce the model which we will use to access the feasibility of our proposed approach to setting up a partially reflective boundary for UPPE. The model describes the propagation of TE electromagnetic waves in a homogeneous medium, that we for convenience assume is a vacuum. We then proceed to set up and solve the eigenvalue problem for the complex transverse wave numbers that define the leaky modes. In this section we also derive a very accurate explicit asymptotic formula for the location of leaky mode wave numbers in the complex plane. In section three we introduce the leaky modes and show that, by using the  technique, well know from the study of resonant states in quantum theory,  of shifting them over to a complex spatial contour outside the transverse computational domain, the leaky modes can be identified with vectors in a vector space of functions on the real line that is endowed with a complex non-Hermitian inner product\cite{David}\cite{Brown}. The leaky modes are orthogonal with respect to this product and we can thus write down generalized Fourier series for any given function based on the orthogonal leaky modes and this non-Hermitian inner product. This solves  the projection problem for our leaky modes.

  We have done extensive numerical experiments using our leaky mode expansions and in section four we presents some examples, and the conclusions we draw from these examples, with regards to their suitability for representing initial data for UPPE. We argue that the leaky mode expansions converge point wise for all sufficiently smooth functions in our space, but that they do not always converge to the function used to generate the expansion. The point wise convergence only becomes problematic in the limit when the index, of the artificial material introduced outside the computational domain, tends towards the same value as the index  inside the domain.  One would expect that problems with the leaky mode expansions would appear in this limit, since if the limit is reached, there is no index difference between the inside and  the outside of the computational domain and leaky modes cease to exist. However, in order to minimize the reflection from the boundary of the computational domain we want to choose the difference between the inside index and the outside, artificial index, as small as possible. It thus becomes a trade off between making it small in order to minimize reflections, and not making it so small that the leaky mode expansions stops giving a good representation of the functions used to generate the expansions. At the end of section four we argue, using a dimensionless quantity that appear from our theory, that there is an acceptable trade off that can be made.
  
  In this paper we do not present a proof that the leaky mode expansion converge to the function used to generate the series. The chief reason for this is that we believe that they never really do converge point wise to the function used to generate them. This is what our numerical results from section four indicated. In section five we present analytical arguments that points to the same conclusion. However, the conjectured lack of point wise convergence to the desired function does not make the leaky mode expansions useless from a more practical point of view. This what we argued in section four where we used a certain dimensionless quantity to specify what we mean by a practical point of view in this context.

\section{The model}
\begin{figure}[t]
  \centering
\captionsetup{width=0.85\textwidth}
\begin{subfigure}{.5\textwidth}
  \centering
  \includegraphics[scale=0.4]{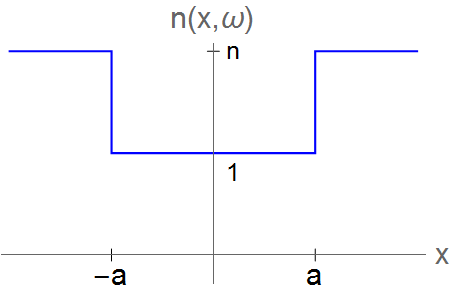}
  \caption{Refractive index $n(x,\omega)$}
  \label{fig1a}
\end{subfigure}%
\begin{subfigure}{.5\textwidth}
  \centering
  \includegraphics[scale=0.4]{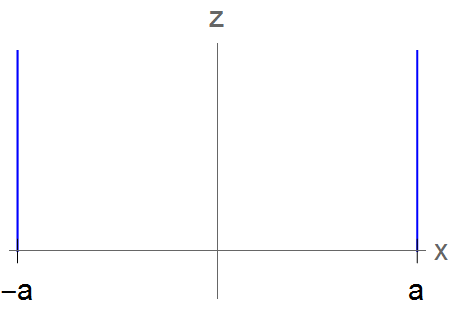}
  \caption{Geometry of the channel}
  \label{fig1b}
\end{subfigure}
\caption{}
\label{fig1}
\end{figure}

In a situation where there are no free charges or currents, Maxwell's equations in the frequency domain can be written in the form
\begin{align}
\nabla\times\textbf{E}&=-i\omega\textbf{B},\label{eq6}\\
\nabla\times\textbf{B}&=i\omega\mu_0\textbf{D},\label{eq5}\\
\nabla\cdot\textbf{E}&=0,\label{eq4}\\
\nabla\cdot\textbf{B}&=0.\label{eq7}
\end{align}
Here, we use the sign convention for the inverse Fourier transform with respect to time that is standard in optics,
\begin{align}
\textbf{E}(\textbf{r},t)=\int_{-\infty}^{\infty}\;d\omega\;\textbf{E}(\textbf{r},\omega)\;e^{-i\omega t}.\label{FourierConvention}
\end{align}
The polarization is a sum of a linear and a non-linear part. The linear part, which is the focus of the current paper, takes  in  frequency domain, for a material that is dispersive and possibly spatially inhomogeneous, the form
\begin{align}
\textbf{P}_{L}(\textbf{r},\omega)=\varepsilon_0\chi(\omega)\textbf{E}(\textbf{r},\omega),
\end{align}
and thus the electric displacement takes the form
\begin{align}
\textbf{D}(\textbf{r},\omega)&=\varepsilon_0\textbf{E}(\textbf{r},\omega)+\textbf{P}(\textbf{r},\omega)=\epsilon_0 n^2(\omega,\textbf{x})\textbf{E}(\textbf{r},\omega)+\textbf{P}_{NL}(\textbf{r},\omega),\label{eq2}
\end{align}
where $n=n(\omega,\textbf{x})$ is the refractive index of the material, defined as usual by the identity
\begin{align}
n^2(\omega,\textbf{x}))=1+\chi(\omega,\textbf{x}).
\end{align}
We will now assume that the spatial inhomogeneity  of the refractive index takes the form of a straight channel oriented along the z-axis of our coordinate system, of uniform width $2a$ in the transverse direction, which is oriented along the x-axis of our coordinate system. The geometry of the channel is illustrated in figure \ref{fig1}.
Consistent with the geometry we assume that the electromagnetic field is transverse electric. Thus we have
\begin{align}
\textbf{E}(\textbf{r},\omega)&=(0,e(x,z,\omega),0), \nonumber\\
\textbf{P}(\textbf{r},\omega)&=(0,p(x,z,\omega),0),\label{eq9}
\end{align}

Using Maxwell's equations we find that  $e(x,z,\omega)$ is a solution to the following model equation
\begin{align}
\partial_{zz}e(x,z,\omega)+\partial_{xx}e(x,z,\omega)+&\left(\frac{\omega}{c}\right)^2(1+\chi(x,\omega))e(x,z,\omega)=p(x,z,\omega),\label{eq10}\\
\nonumber\\
&\Downarrow\text{linearization}\nonumber\\
\nonumber\\
\partial_{zz}e(x,z,\omega)+\partial_{xx}e(x,z,\omega)+&\left(\frac{\omega}{c}\right)^2n^2(x,\omega)e(x,z,\omega)=0.\label{eq11}
\end{align}
In addition to the model equation (\ref{eq10}),  the electric field $e(x,z,\omega)$ must satisfy the conditions
\begin{align}
e(\pm a_-,z,\omega)&=e(\pm a_+,z,\omega),\label{eq12}\\
\partial_x e(\pm a_-,z,\omega)&=\partial_x e(\pm a_+,z,\omega),\label{eq13}
\end{align}
which follows from the electromagnetic interface conditions for transverse electric fields at $x=\pm a$.

The goal is now to  find leaky modes for the linearized equation. These modes can then be used to write  down a  UPPE version of the nonlinear equation (\ref{eq10}), where a leaky mode transform takes the place of the usual transverse Fourier transform. The rest of the paper is focused on constructing the leaky modes and evaluating for which transverse field configurations they form a suitable basis.

\section{Leaky modes}
Leaky modes are solutions to the linearized model equation (\ref{eq11}) that are propagating in the direction of the  the positive z-axis, satisfy the electromagnetic interface conditions (\ref{eq12}),(\ref{eq13}),  and are outgoing at positive and negative infinity. 

Such functions must be of the form
\begin{align}
&e(x,z,\omega)=De^{i\beta z}e^{i\xi x},& x>a,\nonumber\\
&e(x,z,\omega)= e^{i\beta_0z}\left(Be^{i\xi_0x}+Ce^{-i\xi_0x}\right),  & -a<x<a,\nonumber\\
&e(x,z,\omega)=Ae^{i\beta z}e^{-i\xi x},& x<-a,\label{eq14}
\end{align}
where $\beta_0,\xi_0$ and $\beta,\xi$ are the propagation constants and transverse wave numbers  inside and outside the channel, respectively. The propagation constants are determined by the transverse wave numbers by the identities 
\begin{align}
\beta&=\left(\left(\frac{\omega}{c}\right)^2n^2-\xi^2\right)^\frac{1}{2},\nonumber\\
\beta_0&=\left(\left(\frac{\omega}{c}\right)^2-\xi_0^2\right)^\frac{1}{2}.\label{eq16}
\end{align}
\begin{figure}[t]
  \centering
\captionsetup{width=0.85\textwidth}
  \includegraphics[scale=0.25]{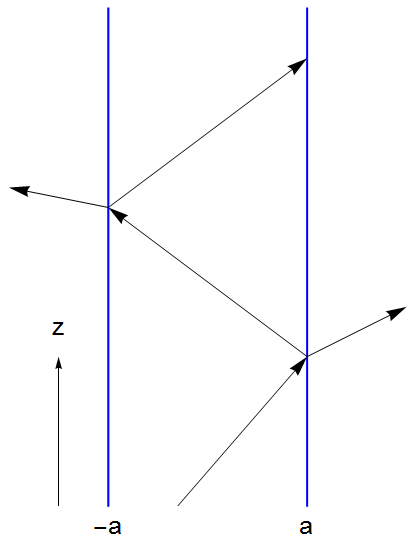}
  \caption{Plane-wave propagation.}
\label{fig2}
\end{figure}
From the physical point of view, the modes represents electromagnetic disturbances that propagate in the direction of the positive z-axis while they are partially reflected and transmitted at the lateral boundaries defining the index channel. This is illustrated in figure \ref{fig2}.

 In order for the functions (\ref{eq14}) to satisfy the electromagnetic boundary conditions (\ref{eq13}), and thus be leaky modes,  the two propagation constants $\beta$ and $\beta_0$ must be equal, which is only true if the following identity holds 
\begin{align}
\xi^2=\left(\frac{\omega}{c}\right)^2(n^2-1)+\xi_0^2.\label{eq17}
\end{align}
 This is Snell’s law. In addition, the following linear algebraic system 
 
\begin{align}
\begin{pmatrix}
 e^{i a\xi } & -e^{-ia \xi _0} & -e^{i a\xi _0} & 0 \\
 -i e^{i a\xi } \xi  & -i e^{-i a\xi _0} \xi _0 & i e^{i a\xi _0} \xi _0 & 0 \\
 0 & e^{i a\xi _0} & e^{-i a\xi _0} & -e^{i a\xi } \\
 0 & i e^{i a\xi _0} \xi _0 & -i e^{-i a\xi _0} \xi _0 & -i e^{i a\xi } \xi  \\
\end{pmatrix}
\begin{pmatrix}
A\\
B\\
C\\
D
\end{pmatrix}=
\begin{pmatrix}
0\\
0\\
0\\
0
\end{pmatrix},
\label{eq18}
\end{align}
 must have a unique solution. This can only happen if the the determinant of the matrix defining the system is zero. One can show that the determinant is zero if and only if the transverse wavenumber satisfy following equation 
\begin{align}
\tan(2a\xi_0)+i\frac{2\xi\xi_0}{\xi^2+\xi_0^2}=0.\label{eq19}
\end{align}
Equation (\ref{eq19}), together with Snell’s law (\ref{eq17}), will determine the dispersion law pertaining to each separate leaky mode. 

 Note that the system of (\ref{eq17})-(\ref{eq19}) has two symmetries connecting solutions. If we denote solutions using the notation $\left\{\left\{\xi,\xi_0\right\},\left(A,B,C,D\right)\right\}$, the two symmetries are of the form
 \begin{align}
 \left\{\left\{\xi,\xi_0\right\},\left(A,B,C,D\right)\right\}&\rightarrow\left\{\left\{\xi,-\xi_0\right\},\left(A,C,B,D\right)\right\},\label{Symmetry1}\\
  \left\{\left\{\xi,\xi_0\right\},\left(A,B,C,D\right)\right\}&\rightarrow\left\{\left\{-\xi^*,-\xi_0^*\right\},\left(A^*,B^*,C^*,D^*\right)\right\}.\label{Symmetry2}
  \end{align}
 Let us start by observing that using the symmetries (\ref{Symmetry1}) and (\ref{Symmetry2}), it is enough to find all the solutions 

\subsection{Dispersion laws}
In this section we will design asymptotic formulas for all solutions $\xi,\xi_0$ to equations (\ref{eq17}),(\ref{eq19}), and thus determine all modes for the system and their respective dispersion laws.

Let us start by observing that $\xi_0=0$ is a solution to equation (\ref{eq19}) and that the corresponding solution  vector to the linear system (\ref{eq17}) is given by 
\begin{align}
\begin{pmatrix}
A\\
B\\
C\\
D
\end{pmatrix}=
\begin{pmatrix}
0\\
-1\\
1\\
0
\end{pmatrix}.\label{eq21}
\end{align}
However, we observe that if we insert $\xi_0=0$ and the vector (\ref{eq21}) into the formula for the modes  (\ref{eq14}) we find that the corresponding mode is identically zero. Thus the solution  $\xi_0=0$ only gives us a trivial mode which can be disregarded when we use the modes for expanding electric field configurations.

\begin{figure}[t]
  \centering
\captionsetup{width=0.85\textwidth}
  \includegraphics[scale=0.4]{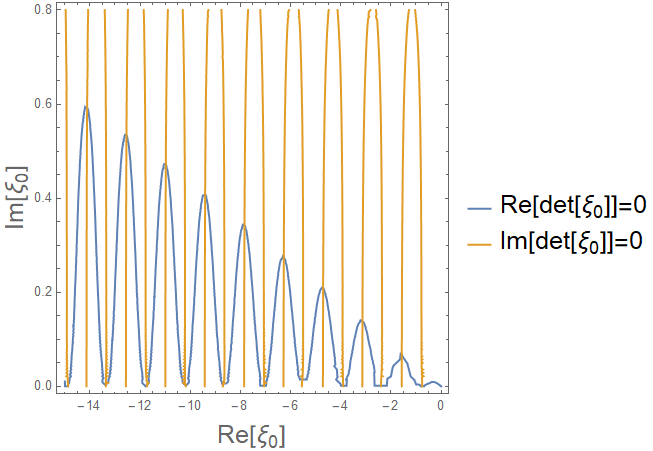}
  \caption{Zero contours of the determinant in the second quadrant. Parameter values  used in this plots were $a=1,\omega/c\approx 1.58153\times 10^7,n=1+10^{-12}$.}
  \label{fig3}
\end{figure}

Observe that because of the symmetries (\ref{Symmetry1}) and (\ref{Symmetry2}), it is sufficient to consider the case when $\xi_0$ is in the second quadrant. Any solution in one of the other quadrants can be generated from a solution in the second quadrant by using the symmetries. In the second quadrant, we can split the system (\ref{eq17}),(\ref{eq19}) into to separate systems depending on which square root  we take when equation (\ref{eq17}) is used to express $\xi$ as a function of $\xi_0$
\begin{align}
\tan(2a\xi_0)&= -i\frac{2\xi_0\sqrt{\alpha+\xi_0^2}}{\xi^2+\xi_0^2},\quad\quad \xi=\sqrt{\alpha+\xi_0^2}\label{case1},\\
\tan(2a\xi_0)&= i\frac{2\xi_0\sqrt{\alpha+\xi_0^2}}{\xi^2+\xi_0^2},\quad\quad \xi=-\sqrt{\alpha+\xi_0^2}\label{case2},
\end{align}

where we have defined $\alpha=(\omega/c)^2(n^2-1)$.  In figure (\ref{fig3}) we display the solutions of the first of the two systems, (\ref{case1}). In the figure, the solutions are defined by the intersection of the zero contours for the real and imaginary part of the equation for $\xi_0$ in (\ref{case1}). There clearly exists an infinite set of solutions, each one corresponding to a distinct mode with its associated dispersion law. A similar plot for the second of the two systems, (\ref{case2}),  gives convincing numerical evidence that it has no solutions in the second quadrant and thus this system does not give us any additional modes  in the second quadrant.
 
It is evident that for  most solutions displayed in figure (\ref{fig3}), the real part strongly dominate the imaginary part.  This fact can be used to find an asymptotic formula for the solutions to equation (\ref{eq19}). 

Assuming that $|\xi_0|\gg\sqrt{\alpha}$, equation  (\ref{eq19}) can be approximated by
\begin{align}
\tan(2a\xi_0)=- i\frac{2\xi_0|\xi_0|\sqrt{1+\frac{\alpha}{\xi_0^2}}}{2\xi_0^2\left(1+\frac{\alpha}{2\xi^2_0}\right)}= i\sqrt{1+\frac{\alpha}{\xi_0^2}}\left(\frac{1}{1+\frac{\alpha}{2\xi^2_0}}\right)\approx i\left(1-\frac{\alpha^2}{8 x^4}\right),\label{eq20}
\end{align}
because  $|\xi_0|\equiv\sqrt{\xi_0^2}$, is equal to $-\xi_0$ when $\xi_0$ is in the second quadrant.

\noindent

Judging from the locations of the zeros in figure (\ref{fig3}),  most of them will be found in regions of the complex plane where $|\text{Re}[\xi_{0}]|\gg|\text{Im}[\xi_{0}]|$. We therefore  write $\xi_{0}= x+iy$, where $|x|\gg|y|$, and use this to simplify  the first equation from  (\ref{eq20})  as follows
\begin{align}
-i\frac{e^{i(2ax+i2ay)}-e^{-i(2ax+i2ay)}}{e^{i(2ax+i2ay)}+e^{-i(2ax+i2ay)}}\approx i\left(1-\frac{\alpha^2}{8 x^4}\right),\nonumber\\
-\frac{e^{i(4ax+i4ay)}-1}{e^{i(4ax+i4ay)}+1}\approx \left(1-\frac{\alpha^2}{8 x^4}\right),\nonumber\\
-\frac{re^{i\theta}-1}{re^{i\theta}+1}\approx\left(1-\frac{\alpha^2}{8 x^4}\right),\label{eq23}
\end{align}
where $r=\text{exp}(-4ay)$ and $\theta=4ax$. We find the real and imaginary part of the left-hand side of equation (\ref{eq23}) to be
\begin{align}
\frac{1-r^2}{r^2+2r\cos\theta+1}-i\frac{2r\sin\theta}{r^2+2r\cos\theta+1}\approx\left(1-\frac{\alpha^2}{8 x^4}\right).\label{eq24}
\end{align}
The right-hand side in equation (\ref{eq24}) is real, so we must have
\begin{align}
\theta&=k\pi\Rightarrow x=\frac{k\pi}{4a},\label{eq25}
\end{align}
where $k$ is a whole number. We know from figure (\ref{fig3}), that for this choice of $\xi$ the determinant has solutions in the $2^{nd}$ quadrant, so $k$ is a negative whole number. Also, we assume that $y>0$. According to equation (\ref{eq25}), $\theta$ is a whole number multiple of $\pi$. Let $k$ be first an odd number $k=2p+1$. Then the imaginary part of the left-hand side in equation (\ref{eq24}) we get
\begin{align}
\frac{1-r^2}{r^2+2r\cos\theta+1}&=\frac{(1-r)(1+r)}{r^2-2r+1}=\frac{1+r}{1-r}.\label{eq26}
\end{align}
Notice that this result is $>1$, while the real part of the right-hand side in equation (\ref{eq24}) is $<1$. This is a contradiction which tells is that $k$ cannot be odd. On the other hand, with $k$ being even, $k=2p$, we get
\begin{align}
\frac{1-r}{1+r}&=1-\frac{\alpha^2}{8x^4},\label{eq27}
\end{align}
where both sides are less than one. Solving equation (\ref{eq27}) for $r$ and consequently for $y$ we get
\begin{align}
r&=\frac{\alpha^2}{16x^4-\alpha^2}\approx\frac{\alpha^2}{16x^4},\nonumber\\
&\Downarrow\nonumber\\
e^{-4ay}&=\frac{\alpha^2}{16x^4},\nonumber\\
y&=-\frac{1}{4a}\text{Log}\left(\frac{\alpha^2}{16x^4}\right)=\frac{1}{4a}\text{Log}\left(\frac{16x^4}{\alpha^2}\right).\label{eq28}
\end{align}
This gives us the approximative solutions to equation (\ref{eq19}) in the $2^{nd}$ quadrant.
\begin{align}
\xi_{0p}&=-\frac{p\pi}{2a}+i\frac{1}{4a}\text{Log}\left(\frac{\pi^4p^4}{\alpha^2a^4}\right),\label{eq29}
\end{align}
where $p=1,2,\cdots$. In order to get an even better approximation, we can write eq. (\ref{eq27}) as
\begin{align}
\frac{1-r}{1+r}&=1-\frac{\alpha^2}{8\xi_0^4},\label{eq30}
\end{align}
where the fourth power on the right-hand side allows us to write $\xi_0$ instead of $x$ , because we work with asymptotic expression under the assumption that $|x|\gg|y|$. This leads to the following iteration scheme for the solutions $\xi_{0p}$
\begin{align}
\xi_{0p}^{n+1}&=-\frac{p\pi}{2a}+i\frac{1}{4a}\text{Log}\left(\frac{16\left(\xi_{0p}^{n}\right)^4}{\alpha^2}\right),\label{eq31}
\end{align}
where $n$ is the index in the recursive formula with  $\xi_{0p}^0=-p\pi/(2a)$. As it turns out, we get a very  good approximation already for  $n=2$ and thus an asymptotic approximation to the solutions of (\ref{eq17}),(\ref{eq19}) in the 2$^{st}$ quadrant is
\begin{align}
\xi_{0p}&=-\frac{p\pi}{2a}+\frac{i}{4a}\text{Log}\left[\frac{16\left(-\frac{p\pi}{2a}+\frac{i}{4a}\ln\left[\frac{p^4\pi^4}{\alpha^2a^4}\right]\right)^4}{\alpha^2}\right].\label{eq32}
\end{align}

Using the symmetries (\ref{Symmetry1}),(\ref{Symmetry2}) we get the formulas for modes residing in the other three quadrants in the form
\begin{align}
\text{1$^{st}$}:\quad \xi_{0p}&=\frac{p\pi}{2a}+\frac{i}{4a}\text{Log}\left[\frac{16\left(\frac{p\pi}{2a}+\frac{i}{4a}\ln\left[\frac{p^4\pi^4}{\alpha^2a^4}\right]\right)^4}{\alpha^2}\right],\label{eq33}\\
\text{3$^{rd}$}:\quad \xi_{0p}&=-\frac{p\pi}{2a}-\frac{i}{4a}\text{Log}\left[\frac{16\left(-\frac{p\pi}{2a}-\frac{i}{4a}\ln\left[\frac{p^4\pi^4}{\alpha^2a^4}\right]\right)^4}{\alpha^2}\right],\label{eq34}\\
\text{4$^{th}$}:\quad \xi_{0p}&=\frac{p\pi}{2a}-\frac{i}{4a}\text{Log}\left[\frac{16\left(\frac{p\pi}{2a}-\frac{i}{4a}\ln\left[\frac{p^4\pi^4}{\alpha^2a^4}\right]\right)^4}{\alpha^2}\right].\label{eq35}
\end{align}
Writing the formula defining $\xi$ in terms of $\xi_0$ from (\ref{case1}) in the form $\xi=\xi'+i\xi''$, it is evident that if $\xi_0$ is in the second quadrant, then $\xi'>0$ and  $\xi''<0$. Using the formula (\ref{eq14}) and our convention for the inverse Fourier transform (\ref{FourierConvention}) we can conclude that the modes in the second quadrant, determined by formula (\ref{eq32}), are outgoing and exponentially growing in the transverse direction. From the formula for the propagation constants (\ref{eq16}) it is also evident that they are decaying in the propagation direction. These are thus leaky modes. In a similar way the modes determined by formula (\ref{eq35}) are also outgoing and decaying in the propagation direction, and thus are also leaky modes. We find however that the  modes determined by formulas (\ref{eq33}) and (\ref{eq34}) are incoming and growing in the propagation direction. These modes are thus not leaky modes, but gaining modes.

Even if we assumed $p$ being large, the formulas for all four quadrants give surprisingly good results, even when $p$ is of order 1. However, it is exactly in this region where the formulas can break down.  Observe that the inner logarithm in formulas (\ref{eq33}-\ref{eq35}) must be positive in order to stay in the same quadrant. Therefore, these formulas  become invalid if
\begin{align}
\frac{p^4\pi^4}{\alpha^2a^4}&\lesssim 1,\nonumber\\
p\lesssim\frac{a\sqrt{\alpha}}{\pi}.\label{eq36}
\end{align}
Let us assume that $\sqrt{\alpha}\gg|\xi_{0p}|$. Applying this assumption to the equation for $\xi_0$ from (\ref{case1}), which determine the leaky modes in the 2$^{st}$ quadrant, gives us 
\begin{align}
\tan(2a\xi_0)= -i2\frac{\xi_0\sqrt{\alpha\left(1+\xi_0^2/\alpha\right)^2}}{\alpha\left(1+2\xi_0^2/\alpha\right)}&\approx -i\frac{2\xi_0}{\sqrt{\alpha}},\nonumber\\
\Downarrow\nonumber\\
\frac{2r\sin\theta}{r^2+2r\cos\theta+1}+i\frac{1-r^2}{r^2+2r\cos\theta+1}&\approx -i\frac{2\xi_0}{\sqrt{\alpha}},\label{eq37}
\end{align}
where $\xi_0=x+iy$ and $r=\text{exp}(-4ay)$, $\theta=4ax$. Using the same approach as before, we find that
\begin{align}
\xi_{0p}&=-\frac{\pi p}{2a}+i\frac{1}{4a}\text{Log}\left[\frac{\sqrt{\alpha}-2\xi_0}{\sqrt{\alpha}+2\xi_0}\right]\approx-\frac{\pi p}{2a}+i\frac{1}{4a}\text{Log}\left[1-\frac{4\xi_0}{\sqrt{\alpha}}\right]\approx-\frac{\pi p}{2a}-i\frac{\xi_0}{a\sqrt{\alpha}}.\label{eq38}
\end{align}
Under the assumed condition $\sqrt{\alpha}\gg|\xi_{0p}|$, the second term in equation (\ref{eq38}) is a small correction to the first term. This allows us to look at equation (\ref{eq38}) as a recursion formula for the solution $\xi_{0p}$. Starting from the leading term $\xi_{0p}^0=-\pi p/(2a)$, we obtain from the first iteration
\begin{align}
\xi_{0p}&\approx-\frac{\pi p}{2a}+i\frac{1}{a\sqrt{\alpha}}\frac{\pi p}{2a}=-\frac{\pi p}{2a}+i\frac{\pi p}{2a^2\sqrt{\alpha}}.\label{eq39}
\end{align}
We have thus obtained a different asymptotic formula for the solutions $\xi_{0p}$ in the 2$^{st}$ quadrant, a formula where we know that the imaginary part is a small correction to the real part,  $\pi p/(2a)\gg\pi p/(2a^2\sqrt{\alpha})$,  or equivalently, $p\ll a\sqrt{\alpha}$.  This condition implies that condition (\ref{eq36}) holds. Thus, we can conclude that the asymptotic formula (\ref{eq39}) holds exactly when the asymptotic formula (\ref{eq33}) breaks down. Formulas similar to (\ref{eq39}) can be derived for the other quadrants.

Some of the leaky modes are paraxial whereas others are not.  In order to be more precise about which modes are paraxial, note that the propagation vector for the light beam is of the form $(\xi_0,\beta_0)$. This is clear from equation (\ref{eq14}). This allows us to calculate the propagation angle of the beam with respect to the $z$-axis. This angle is 
\begin{align}
\theta_p=\tan^{-1}\left(\frac{\text{Re}\left[\xi_{0p}\right]}{\text{Re}\left[\beta\left(\xi_{0p}\right)\right]}\right).\label{eq40}
\end{align}
  Clearly, for each $\xi_{0p}$, we get a different angle. In order for a mode to be paraxial, the angle $\theta_p$ must be small, and this holds only if 
  \begin{align}
  \text{Re}\left[\xi_{0p}\right]&\ll \text{Re}\left[\beta\left(\xi_{0p}\right)\right],\nonumber\\
  &\Updownarrow\nonumber\\
  p&\ll \frac{a \sqrt{2}\omega}{\pi c},\label{ParaxialModes}
  \end{align}
where we have used the fact to leading order  $\xi_{0p}\approx-\frac{\pi p}{2 a}$. Formula (\ref{ParaxialModes}) determine which leaky modes are paraxial.

We now investigate if there are zeros in parts of the complex $\xi_0$-plane that are not covered by the asymptotic formulas we have found so far. We will focus on the second quadrant, the other quadrants can be treated in a similar way with corresponding results. These investigations are necessary, because the exponential smallness of the equation determining $\xi_0$, in the part of the second quadrant well away from the real axis,  makes a direct numerical search for solutions, like the one in figure \ref{fig3}, very challenging.

Let us first look for zeros in  the part of the second quadrant where  $y=\text{Im}[\xi_{0}]$ is much larger than $x=\text{Re}[\xi_{0}]$ and $\text{Im}[\xi_{0}]\gg 1\gg\alpha$.  Under these conditions on $\xi_{0}$ we have $\tan(2a\xi_0)\approx i$ and equation  for $\xi_0$ in (\ref{case1}) takes the simplified form
\begin{align}
i&=i\left(1-\frac{\alpha^2}{8\xi_0^4}+\frac{\alpha^3}{8\xi_0^6}\right),\nonumber\\
0&\approx-\frac{\alpha^2}{8y^4}+\frac{\alpha^3}{8y^6},\nonumber\\
y&=\sqrt{\alpha},\label{eq42.1}
\end{align}
where we included the next term of the Taylor series for the right-hand side in equation (\ref{case1}). We thus end up with a solution $\xi_0=i\sqrt{\alpha}$ that contradicting the assumptions imposed on $\xi_{0}$.  Hence, no zeros can exist in this part of the second quadrant.

Let us next look at the region where $y=\text{Im}[\xi_{0}]\sim \text{Re}[\xi_{0}]=x$ and $y,x\gg 1\gg\alpha$. At this point, observe that the left hand side of  equation (\ref{eq19}) is not actually equal to the determinant of the system(\ref{eq18}).  Imposing equation  (\ref{eq19}) only  implies that the determinant is zero.  If we rather equates the full determinant of (\ref{eq18}) to zero we get
\begin{align}
\left(e^{4ia\xi_0}-1\right)\xi_0^2-2\left(e^{4ia\xi_0}+1\right)\xi\xi_0+\left(e^{4ia\xi_0}-1\right)\xi^2&=0,\nonumber\\
-(\xi_0+\xi)^2+e^{4ia\xi_0}(\xi_0-\xi)^2&=0,\nonumber\\
e^{4ia\xi_0}&=\frac{(\xi_0+\xi)^2}{(\xi_0-\xi)^2}.\label{eq42.2}
\end{align}
The above equation can be simplified using the assumptions $|\xi_0|\gg\alpha$. Writing $\xi_0=-x+iy$, where $x,y>0$ we get
\begin{align}
\frac{\alpha^2}{16\xi_0^4}&=0,\nonumber\\
\frac{(-x-iy)^4}{16(x^2+y^2)^4}&=0,\nonumber\\
&\Downarrow\nonumber\\
\text{Re:}\quad x^4-6x^2y^2+y^4&=0,\label{eq42.3}\\
\text{Im:}\quad\quad\quad\;\; x^3y-xy^3&=0.\label{eq42.4}
\end{align}
The only possible solutions to equation (\ref{eq42.3}) are $x=y(\sqrt{2}-1),x=y(\sqrt{2}+1)$. Substituting these solutions into (\ref{eq42.4}), and solving for $y$, we find in both cases t $y=0$. This contradict our assumptions and thus there are no solutions in this region of the second quadrant either. We have now covered all possible regions of the second quadrant and thus conclude that there are no other zeros of the determinant, and thus leaky modes, than the ones we have already found and that is covered by our asymptotic formulas.

There is however one remaining issue related to the zeros, and thus leaky modes, that needs to be discussed. As we have already noted, the asymptotic formulas for the zeros, which, by design, are expected to be accurate only in the limit when the index $p$ is very large,  in fact works surprisingly well even for $p$ as small as 2. However, the very first zero, the one corresponding to $p=1$, is never very accurate. The first zero also behave differently when the parameter $\alpha$ is varied. Recall that the value of this parameter is proportional to the size of the index step defining the channel where the waves will be propagating. We are interested in minimizing reflections from the edges of the channel and therefore would want to make the index step, and hence the parameter $\alpha$, as small as possible. When we let alpha decrease, we observe that all the zeros in the second quadrant, except the first, move slowly up, and even more slowly towards the imaginary axis. This behavior is to be expected from of the logarithmic dependence of the imaginary part of the zero on the parameter $\alpha$. The first zero approaches the imaginary axis at at fast rate when $\alpha$ is decreased, and for a finite value of $\alpha=\alpha_c$,  it simply vanishes. For $\alpha<\alpha_c$ we observe that for one value of the index $p$, the formula (\ref{eq33}) indicate the presence of a double zero. The index for which this occurs increase when $\alpha$ keeps decreasing towards zero. These double zeros are however spurious, careful  numerical investigations show that there are no double zeros. However, this abrupt change in the prediction derived from formula (\ref{eq33}),  when $\alpha$ vary smoothly, alerted us to the possibility that the root cause to why our  formula predicted  both the vanishing of the first zero and the existence of double zeros, is the crossing of a branch cut. Observe that the argument of the logarithm in formula (\ref{eq33}) is an expression with complex values, so there is indeed a branch cut implied by the formula and thus the argument crossing this branch cut when $\alpha$ vary smoothly is a real possibility.

 Formula  (\ref{eq32}) can be written in the form
\begin{align}
\xi_{0p}&=-\frac{p\pi}{2a}+\frac{i}{4a}\text{Log}\left[z\right],\nonumber\\
z&=\frac{16\left(-\frac{p\pi}{2a}+\frac{i}{4a}\ln\left[\frac{p^4\pi^4}{\alpha^2a^4}\right]\right)^4}{\alpha^2}.\label{eq96}
\end{align}
We use the standard branch of the logarithm in our calculations, and thus there is a branch cut along the negative real axis.  The real part of $\xi_{0p}$ is negative, so we will have a crossing of the branch cut whenever the imaginary part of $z$ vanish. Expanding the polynomial expression defining $z$ in equation (\ref{eq96}), and taking the imaginary part, we find that there is   a crossing of the  branch cut whenever $\alpha$ solves the equation
\begin{align}
\frac{2p\pi}{a^4}\ln^3\left(\frac{p\pi}{a\sqrt{\alpha}}\right)&=\frac{p^3\pi^3}{2a^4}\ln\left(\frac{p\pi}{a\sqrt{\alpha}}\right).\label{eq97}
\end{align}
Denoting $x=\ln\left(\frac{p\pi}{a\sqrt{\alpha}}\right)$ we have
\begin{align}
x(4x^2-p^2\pi^2)&=0.\label{eq98}
\end{align}
Solutions to eq. (\ref{eq98}) are $x=0,\pm p\pi/2$. Investigating all three solutions we find that the one we are looking for is $+p\pi/2$ which yields
\begin{align}
\alpha_p&=\frac{p^2\pi^2}{a^2}\text{exp}(-p\pi).\label{eq99}
\end{align}
Further numerical investigations show that $\alpha_c=\alpha_1$ and that $\alpha_p$ for $p>1$ correspond to the values of $\alpha$ where formula (\ref{eq33}) predicts a double zero for the value of the index equal to $p$. The impact of the disappearing of the first zero on our leaky mode expansions will be discussed later, at the end of section four.

\subsection{Mode shapes, normalization and projection}
For $\{\xi,\xi_0\}$, solving equations (\ref{eq17}),(\ref{eq19}) with $\xi_0$ in the second quadrant and $\xi$ in the fourth quadrant, this is the case specified in (\ref{case1}), we have a leaky mode whose formula which, according to (\ref{eq14}), is given by
\begin{align}
 u^-_p(x)&=\left\{
\begin{array}{cc}
De^{i\xi_p x}, & x>a\\
 Be^{i\xi_{0p}x}+Ce^{-i\xi_{0p}x}, & -a<x<a\\
 Ae^{-i\xi_p x}, & x<-a
\end{array}\right., \quad \xi_p=\left(\alpha+\left(\xi_{0p}\right)^2\right)^{1/2}.\label{eq44}
\end{align}
Using the symmetries (\ref{Symmetry1}),(\ref{Symmetry2}) we can conclude that there is a corresponding incoming, gaining mode, in the first quadrant whose formula is given by 
\begin{align}
 u^+_p(x)&=\left\{
\begin{array}{cc}
D^*e^{i\xi_p x}, & x>a\\
B^*e^{-i\xi^*_{0p}x}+C^*e^{i\xi^*_{0p}x}, & -a<x<a\\
A^*e^{-i\xi_p x}, & x<-a
\end{array}\right., \quad  \xi_p=-(\alpha+\xi^{*2}_{0p})^{1/2}.\label{eq43}
\end{align}
  
\begin{figure}[t]
  \centering
\captionsetup{width=0.85\textwidth}
  \includegraphics[scale=0.6]{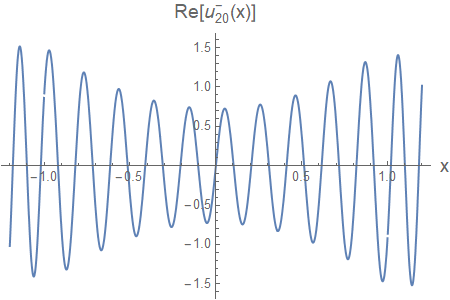}
  \caption{Outgoing mode corresponding to the index $p=20$. Parameters used in this plots were $a=1,\omega/c\approx 1.58153\times 10^7,n=1+10^{-12}$.}
\label{fig6}
\end{figure}

Observe that we have $(u_p^+)^*=u_p^-$.  In figure (\ref{fig6}) we see an outgoing mode corresponding to the index $p=20$. The mode is evidently exponentially growing in $x$. This holds true for all modes, both incoming and outgoing.

Since the modes are exponentially growing in $x$, they are clearly not normalizable. We can make the modes  normalizable by analytically continuing them into a complexified spatial domain, and restricting the analytically continued modes to carefully chosen complex contours. The contours will be different depending on whether the modes are incoming or outgoing. The contours we will be using are of the form 
\begin{align}
z^+(x)=\left\{
\begin{array}{cc}
a-i(x-a), & x>a\\
x, & |x|<a\\
-a-i(x+a), & x<-a
\end{array}\right.,\label{eq45}\\
z^-(x)=\left\{
\begin{array}{cc}
a+i(x-a), & x<a\\
x, & |x|<a\\
-a+i(x+a), & x<-a
\end{array}\right.,\label{eq46}
\end{align}
where $z^+$ is used for the incoming modes and $z^-$ is used for the outgoing modes.

\begin{figure}[t]
  \centering
\captionsetup{width=0.85\textwidth}
  \includegraphics[scale=0.6]{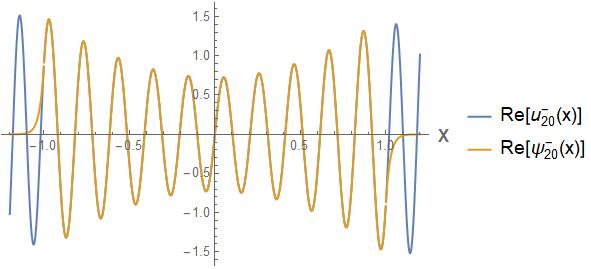}
  \caption{Outgoing mode and its complexified version corresponding to the index $p=20$. Parameters used in this plot were $a=1,\omega/c\approx 1.58153\times 10^7,n=1+10^{-12}$.}
\label{fig7}
\end{figure}

\noindent
Evaluating eq. (\ref{eq43}),(\ref{eq44}) on these contours we find that they exponentially decay in both directions on the real axis. Define functions $\psi^+_p(x)$ and $\psi^-_p(x)$ on the positive real axis as
\begin{align}
\psi^+_p(x)&=u^+_p\left(z^+(x)\right),\label{eq47}\\
\psi^-_p(x)&=u^-_p\left(z^-(x)\right).\label{eq48}
\end{align}
The formulas for these functions are
\begin{align}
 \psi^-_p(x)&=\left\{
\begin{array}{cc}
De^{i\xi_p a}e^{-\xi_p(x-a)}, & x>a\\
Be^{i\xi_{0p}x}+Ce^{-i\xi_{0p}x}, & |x|<a\\
Ae^{i\xi_p a}e^{+\xi_p(x+a)}, & x<-a
\end{array}\right.,\quad\xi_p=(\alpha+\xi_{0p}^2)^{1/2},\label{eq49}\\
\psi^+_p(x)&=\left\{
\begin{array}{cc}
D^*e^{i\xi_p a}e^{\xi_p(x-a)}, & x>a\\
B^*e^{-i\xi_{0p}^*x}+C^*e^{i\xi_{0p}^*x}, & |x|<a\\
A^*e^{i\xi_p a}e^{-\xi_p(x+a)}, & x<-a
\end{array}\right.,\quad \xi_p=-\left(\alpha+\left(\xi_{0p}^*\right)^2\right)^{1/2}.\label{eq50}
\end{align}
Note that for these complexified modes we also have the relation $\left(\psi^+_p(x)\right)^*=\psi^-_p(x)$. Figure (\ref{fig7}) shows the functions $u^-_{20}(x)$ and $\psi^-_{20}(x)$ in the same picture.  We observe that  $\psi^-_{20}(x)$ decay exponentially outside the channel, which is confined to the interval $[-a,a]$. Also note that the complexified modes are not continuously differentiable at the points $x=\pm a$. This is because we restricted the analytically continued modes to a contour that is singular at $x=\pm a$. We made this choice  in order to get fastest possible decay of the complexified modes and the simplest possible expressions for certain key  differential operators acting on the modes. 

Using the analyticity of the complexified modes at the two points $z=\pm a$ ,and the formulas for the two singular contours (\ref{eq46}), it is easy to verify that the following boundary conditions holds for $\psi^+_p(x)$ and $\psi_p^-(x)$ at the two points $x=\pm a$
\begin{align}
\psi^+_p(\pm a^-,\omega)&=\psi^+_p(\pm a^+,\omega), & \psi^-_p(\pm a^-,\omega)&=\psi^-_p(\pm a^+,\omega),\nonumber\\
\partial_x\psi^+_p(-a^-,\omega)&=-i\partial_x\psi^+_p(-a^+,\omega), & \partial_x\psi^-_p(-a^-,\omega)&=i\partial_x\psi^-_p(-a^+,\omega),\nonumber\\
\partial_x\psi^+_p(a^-,\omega)&=i\partial_x\psi^+_p(a^+,\omega), & \partial_x\psi^-_p(a^-,\omega)&=-i\partial_x\psi^-_p(a^+,\omega).\label{eq51}
\end{align}
This fact tells us that complexified modes $\psi^+,\psi^-$ belong to two different spaces of functions, $V^+$ and $V^-$. Here $V^-$ is the space of smooth functions on real line which satisfies the boundary conditions for $\psi_p^-(x)$  (\ref{eq51}), and similarly for $V^+$.  We evidently have 
\begin{align}
\left\{\psi^+_p(x)\right\}_{p=1}^\infty&\subset V^+, & \left\{\psi^-_p(x)\right\}_{p=1}^\infty&\subset V^-.\label{eq52}
\end{align}

\noindent
It is easy to verify that the complexified modes are in fact eigenfunctions to the differential operator 
\begin{align}
\mathcal{L}_x&=\left\{
\begin{array}{cc}
\partial_{xx}+\left(\frac{\omega}{c}\right)^2, & |x|<a\\
 & \\
-\partial_{xx}+\left(\frac{\omega}{c}\right)^2(n^2-1), & |x|>a
\end{array}\right..\label{eq55}
\end{align}
We have
\begin{align}
\mathcal{L}_x\psi^-(x)&=\lambda_p\psi^-_p(x),\quad \lambda_p=\left(\left(\frac{\omega}{c}\right)^2-\xi_{0p}^2\right)^\frac{1}{2},\label{eq53}\\
\mathcal{L}_x\psi^+(x)&=\mu_p\psi^+_p(x),\quad \mu_p=\lambda_p^*.\label{eq54}
\end{align}

In order use the complexified modes as a tool for expanding functions  in $V^-$, functions  that are in the span of $\left\{\psi^-_p(x)\right\}_{p=1}^\infty$, we need an inner product on the space. Furthermore, with respect to this inner product the leaky modes must be normalizable and orthogonal. Orthogonality would be assured if the operator $\mathcal{L}_x$, defined in (\ref{eq55}), is self-adjoint with respect to the chosen inner product. This, however, seems like an impossible task, since we know that the eigenvalues $\lambda_p$, defined  in (\ref{eq54}), are in fact complex.

Nevertheless, an inner product that satisfy all the requirements can be constructed. In order to do this, note 
that for  any contour $\mathcal{C}$ in the complex plane we can define a complex values scalar product on the space of functions analytic in an open set containing the contour
\begin{align}
(\Phi,\Psi)&=\int_\mathcal{C}\Phi(z)\overline{\Psi}(z)\mathrm{d}z\in\mathbb{C},\label{eq56}
\end{align}
where $\overline{\Psi}(z)$ is an analytic function defined by $\overline{\Psi}(z)=\Psi^*\left(z^*\right)$. Applying this definition of scalar product of analytic functions to the contour $z^-$, we get the following complex valued scalar product on the space $V^-$, defined for any pair of function $\psi,\phi\in V^-$ by the expression
\begin{align}
\left(\psi,\phi\right)^-&=i\int_{-\infty}^{-a}\psi(x)\phi(x)\mathrm{d}x+\int_{-a}^{a}\psi(x)\phi(x)\mathrm{d}x+i\int_{a}^{\infty}\psi(x)\phi(x)\mathrm{d}x.\label{eq58}
\end{align}
It is now straight forward to show that the differential operator $\mathcal{L}_x$ is self-adjoint with respect to the inner product (\ref{eq58})  on the space $V^-$. The orthogonality of the leaky modes then follows by the familiar argument

\begin{align}
(\lambda_p-\lambda_q)\left(\psi^-_p,\psi^-_q\right)^-&=\left(\lambda_p\psi^-_p,\psi^-_q\right)^--\left(\psi^-_p,\lambda_q\psi^-_q\right)^-\nonumber\\
&=\left(\mathcal{L}_x\psi^-_p,\psi^-_q\right)^--\left(\psi^-_p,\mathcal{L}_x\psi^-_q\right)^-\nonumber\\
&=\left(\mathcal{L}_x\psi^-_p,\psi^-_q\right)^--\left(\mathcal{L}_x\psi^-_p,\psi^-_q\right)^-=0,\nonumber\\
&\Downarrow\nonumber\\
\left(\psi^-_p,\psi^-_q\right)^-&=0.\label{eq61}
\end{align}

Any function in $f\in V^-$ which is in the span of the leaky modes $\left\{\psi^-_p(x)\right\}_{p=1}^\infty$ can now be expanded in terms of a generalized Fourier series of the form
\begin{align}
f(x)\in V^-\Rightarrow f(x)&=\sum_{p=1}^\infty\frac{\left(f(x),\psi^-_p\right)^-}{\left(\psi^-_p,\psi^-_p\right)^-}\psi^-_p(x).\label{eq63}
\end{align}

In a similar way an inner product can be introduced on the space of gaining modes $V^+$
\begin{align}
\left(\psi,\phi\right)^-&=-i\int_{-\infty}^{-a}\psi(x)\phi(x)\mathrm{d}x+\int_{-a}^{a}\psi(x)\phi(x)\mathrm{d}x-i\int_{a}^{\infty}\psi(x)\phi(x)\mathrm{d}x,\label{eq58.1}
\end{align}
which can be used to expand gaining modes in a generalized Fourier series of the form 
\begin{align}
f(x)\in V^+\Rightarrow f(x)&=\sum_{p=1}^\infty\frac{\left(f(x),\psi^+_p\right)^+}{\left(\psi^+_p,\psi^+_p\right)^+}\psi^+_p(x).\label{eq62}
\end{align}
Observe that the boundary conditions (\ref{eq51})  implies that $\phi\in V^-\Leftrightarrow\psi^*\in V^+$. Thus the complex conjugate maps between these two spaces. In a similar way, the complex conjugate maps between the inner products on the two spaces
\begin{align}
(\psi,\phi)^{-*}=(\psi^*,\phi^*)^+.
\end{align}
The spaces of leaky modes $V^-$ and gaining modes $V^+$  are not only linear spaces, but also complex algebras. This holds because products of functions preserve the boundary conditions at $x=\pm a$. For any pair of functions in $\psi,\phi\in V^-$ we have for example

\begin{align}
\partial_x(\psi\phi)(a^-)&=(\partial_x\psi)\phi(a^-)+\psi(\partial_x\phi)(a^-)\nonumber\\
&=-i(\partial_x\psi)\phi(a^+)-i\psi(\partial_x\phi)(a^+)\nonumber\\
&=-i((\partial_x\psi)\phi+\psi(\partial_x\phi))(a^+)\nonumber\\
&=-i\partial_x(\psi\phi)(a^+),\nonumber\\
\partial_x(\psi\phi)(a^-)&=-i\partial_x(\psi\phi)(a^+).\nonumber
\end{align}
Thus,  we can conclude that $(\psi\phi)\in V^-$. 

\section{Numerical results}
In this paper we will not present a formal proof specifying precisely which space of functions are in the span of the set of leaky modes, and thus for which space of functions the expansions (\ref{eq63}) converge point wise. We will however present some arguments in section five that addresses the question of convergence of the leaky mode expansions (\ref{eq63}).

In this section we present some numerical tests of the leaky mode expansions that will indicate strongly that they are indeed useful for the optical beam propagation context we have designed them for.  In order for the leaky mode expansions to be useful for modelling (semi) transparent computational boundaries for UPPE there are two conditions that must be met. 

\begin{figure}[t]
  \centering
\captionsetup{width=0.85\textwidth}
\begin{subfigure}{.5\textwidth}
  \centering
  \includegraphics[scale=0.5]{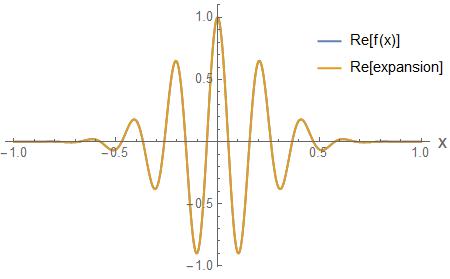}
  \caption{Real part of a gaussian wavepacket and its\\leaky mode expansion.}
  \label{fig8a}
\end{subfigure}%
\begin{subfigure}{.5\textwidth}
  \centering
  \includegraphics[scale=0.5]{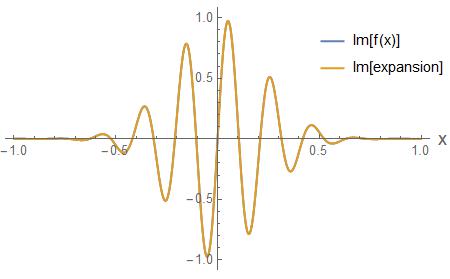}
  \caption{Imaginary part of a gaussian wavepacket\\ and its leaky mode expansion.}
  \label{fig8b}
\end{subfigure}
\caption{A gaussian wavepacket $f(x)=\text{exp}(-mx^2)\text{exp}(ikx)$, where $m=10,k=30$ and its leaky mode expansion using 30 outgoing terms. The parameters used in this expansion were $a=1,\omega/c\approx 1.58153\times 10^7,n=1+10^{-12}$.}
\label{fig8}
\end{figure}

Firstly, physically reasonable initial data must be in the span of the leaky modes. Second, products of functions in the span must also be in the span. In figure (\ref{fig8}) we display  an expansion of a Gaussian wave packet  using only 30 terms in the leaky mode expansion (\ref{eq63}). In these plots, the refractive index outside the slab is $n=1+10^{-12}$. As we can see, the Gaussian wave packet and its  leaky mode expansion are indistinguishable.

Secondly, since UPPE must be able to handle nonlinear interactions, products of functions in the span must also be in the span. In order to investigate this  we expanded squares of the leaky modes, namely, $\left(\psi^-_{11}(x)\right)^2$ and $\left(\psi^-_{50}(x)\right)^2$. The first propagate at an angle of  $2^\circ$ with respect to the $z$-axis, while the second one propagate at an angle of $9^\circ$. The results are clearly very satisfying. Note that for  the second mode we needed more terms in the leaky mode expansion because of its highly oscillatory nature.

\begin{figure}[t]
  \centering
\captionsetup{width=0.85\textwidth}
\begin{subfigure}{.5\textwidth}
  \centering
  \includegraphics[scale=0.5]{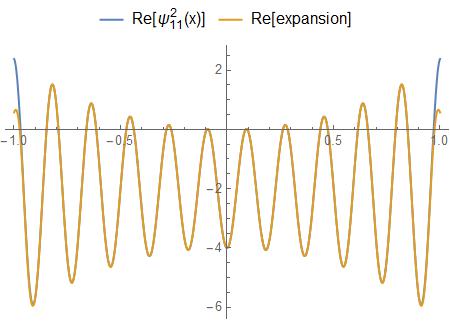}
  \caption{Real part of the square of outgoing mode\\ corresponding to $p=11$ and its leaky mode\\ expansion using 60 outgoing terms.}
  \label{fig9a}
\end{subfigure}%
\begin{subfigure}{.5\textwidth}
  \centering
  \includegraphics[scale=0.5]{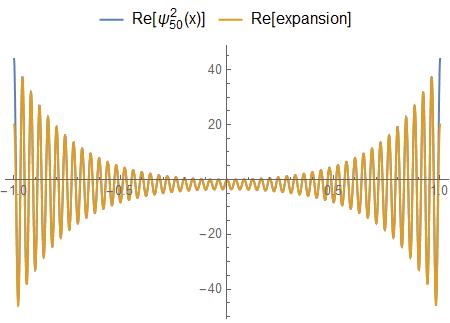}
  \caption{Real part of the square of outgoing mode\\ corresponding to $p=50$ and its leaky mode\\ expansion using 200 outgoing terms.}
  \label{fig9b}
\end{subfigure}
\caption{Two squares of outgoing modes corresponding to $p=11$ and $p=50$ and their leaky mode expansion using 60, resp. 200 outgoing terms. The parameters used in this expansion were $a=1,\omega/c\approx 1.58153\times 10^7,n=1+10^{-12}$.}
\label{fig9}
\end{figure}
In this paper we are not going to implement our leaky mode expansions in a fully nonlinear UPPE propagation algorithm. Before this can be done,  more work has to be put into ensuring  the accuracy and efficiency of  the transformation from a function to its leaky mode expansion and back again. Here we will show a linear propagation example,  where we compare the approach using the leaky modes,  to one using regular Fourier modes, which corresponds to imposing perfectly reflecting boundary condition at at $x=\pm a$. Both are compared to the exact, infinite domain solution,  which, for any $z$,  can be approximated arbitrarily well by using a regular Fourier series on a much larger transverse domain.
To appreciate how well our leaky modes expansion does, we demonstrate a numerical experiment where model a CW Gaussian beam propagation using Fourier expansion for finite as well as for infinite domain and compare it to the leaky modes expansion. 

For the regular Fourier solution with perfectly reflecting boundary conditions at $x=\pm a$ we have 
\begin{align}
e(x,z)&=\sum_{k=-\infty}^\infty \phi_k(x)\text{exp}(i\beta_k z),\label{eq65}\\
\phi_k(x)&=\left\{
\begin{array}{cc}
A_k\cos\left(\pi k x/(2a)\right) & k=2n-1\\
B_k\sin\left(\pi k x/(2a)\right) & k=2n\\
\end{array}\right.,\label{eq66}
\end{align}
where $\beta_k=\left(\left(\pi k x/(2a)\right)^2-(\omega/x)^2\right)^{1/2}$, and where the domain is $x\in(-a,a), z\in(0,\infty)$. As the boundary required by UPPE at $z=0$, we use the Gaussian $e(x,0)=f(x)=\text{exp}(-cx^2)$ for some parameter $c>0$. Notice that the expansion functions in (\ref{eq66}) form an orthogonal set 
\begin{align}
\int_{-a}^a\phi_k(x)\phi_l(x)\mathrm{d}x&=\left\{
\begin{array}{cc}
a & k=l\\
0 & k\neq l
\end{array}\right.,\label{eq67}
\end{align}
for $k,l$ both being either even or odd. Imposing the Gaussian as a boundary condition at $z=0$ determine the coefficients (\ref{eq66}), of the expansion (\ref{eq65})
\begin{align}
A_k&=\frac{1}{a}\int_{-a}^a f(x)\cos\left(\frac{\pi kx}{2a}\right)\mathrm{d}x,\quad k=2n-1,\label{eq68}\\
B_k&=\frac{1}{a}\int_{-a}^a f(x)\sin\left(\frac{\pi kx}{2a}\right)\mathrm{d}x,\quad k=2n.\label{eq69}
\end{align}
As indicated earlier, we also express the infinite domain solution using Fourier modes, now on a larger domain, say $3a$. The exact solution and this numerical solution will not deviate until the diffracting Gaussian hit the boundary of the extended domain. 

\begin{figure}[h!]
  \centering
\captionsetup{width=0.85\textwidth}
\begin{subfigure}{.5\textwidth}
  \centering
  \includegraphics[scale=0.4]{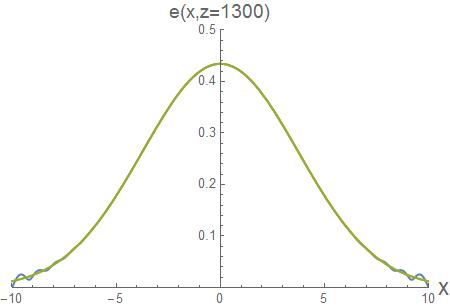}
  \caption{}
  \label{fig10a}
\end{subfigure}%
\begin{subfigure}{.5\textwidth}
  \centering
  \includegraphics[scale=0.4]{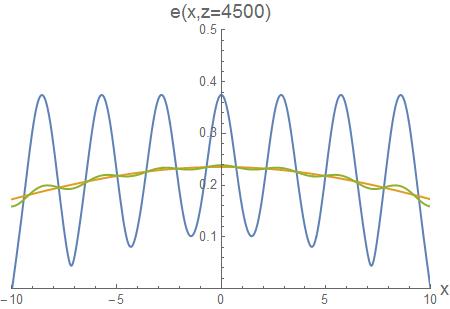}
  \caption{}
  \label{fig10b}
\end{subfigure}
\caption{Comparing different solutions to eq. (\ref{eq11}) using Fourier method in a finite (blue) and infinite domain (orange) and leaky modes (green).  For all methods 100 terms in the expansions were used. Parameters used in this plots were $a=1,\omega/c\approx 1.58153\times 10^7,n=1+10^{-15}$.}
\label{fig10}
\end{figure}

Figure (\ref{fig10}) depicts the solutions for all three approaches. The width of the domain for the finite Fourier method as well as the slab for the leaky modes is $a=10$ and the refractive index outside the slab is $n=1+10^{-15}$ in the optical regime. On (\ref{fig10a}) we see that the leaky modes solution and infinite domain Fourier overlap perfectly, but approximately at this point ($z\approx1300$), the wave hits the boundary of the slab and the finite Fourier solution starts to deviate from the other two solutions, as expected. Propagating the wave further in the slab, we observe that around $z\approx 4500$ the leaky modes solution starts to deviate from the infinite domain Fourier solution. Up to this moment, the leaky modes and infinite Fourier solutions were very close to each other. We can therefore say, that with the leaky modes method we were able to propagate the wave approximately $3.4$ times longer than with finite Fourier method. The reason why the leaky modes eventually collapsed is, that the slab is not perfectly transparent. In other words, the difference in the refractive indices for the slab and the outside domain is non-zero. This leads to reflections that gradually build up as the wave propagates in $z$ causing it to  interfere with itself.

\begin{figure}
  \centering
\captionsetup{width=0.85\textwidth}
  \includegraphics[scale=0.6]{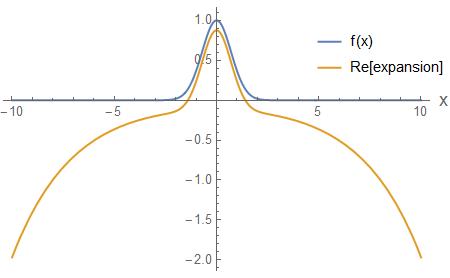}
  \caption{The test function in thi figure is $f(x)=\text{exp}(-x^2)$. Parameters used in this expansion were $a=10,\omega/c\approx 1.58153\times 10^7,n=1+10^{-18}$.}
\label{fig11}
\end{figure}

To be able to propagate the wave using leaky modes even further, we could lower the index $n$ even more, to say $n=1+10^{-18}$. However, it turns out that here we come across some serious issues. Let us first look at an expansions for a Gaussian function using $n=1+10^{-18}$. Looking at figure (\ref{fig11}) we see that the width of the domain is the same as in the propagation example, however we made refractive index $n$ outside the slab closer to 1. The number of terms used in this expansion was 200. The badness of this expansion suggests that one should use perhaps more terms to make it better. But the truth is, the expansion does not change with more terms. Thus, the numerics  indicate that the leaky mode expansion for the Gaussian  does converge point wise, but unfortunately to some other function than the target Gaussian. There are two possible explanations for  what happens here. The first is  that the series actually diverges, but so slowly that we cannot detect it numerically. The second is that the series does converge point wise, but not to the function used to generate it.

Extensive numerical investigations, using very high numerical precision,  leads us to conjecture that it is the second explanation that is correct. In fact, we suspect that the leaky mode {\it never} converge point wise  to the function used to generate it. We will look more into these issues in the next section using asymptotic methods. Here we just note that even though we very likely do not have point wise convergence for the leaky mode expansion, the expansion is nevertheless for the task it was designed for. We find that the deviation between a function and its leaky mode expansion  is only noticeable when the dimensionless number
\begin{align}
\eta=a^2\left(\frac{\omega}{c}\right)^2(n^2-1),\label{eq69.1}
\end{align}
is not too small. For the series to give a, practically speaking, faithful representation  of functions,  we need at least $\eta\gtrsim 10^{-4}$. We find that for a pulse in the optical regime ($\omega/c\approx 1.58153\times 10^7$) we have 
a good representation of Gaussian initial data if
\begin{align}
&a=10^{-1}\text{ m},\quad n\geq 1+10^{-13},\nonumber\\
&a=10^{-2}\text{ m},\quad n\geq 1+10^{-11},\nonumber\\
&a=10^{-3}\text{ m},\quad n\geq 1+10^{-9},\nonumber\\
&a=10^{-4}\text{ m},\quad n\geq 1+10^{-7}.\nonumber
\end{align}
An important  requirement for using the leaky mode expansion is that the main part of the pulse, where the bulk of  the nonlinear interactions takes place, is well inside the domain $[-a,a]$. The choices for the transverse width $a$ of the domain in the above list are chosen because they corresponds to actual dimensions used in high energy, long distance, propagation of optical pulses in air, using the UPPE code developed at the Center for Mathematical Sciences at the University of Arizona.

In the previous chapter we mentioned a phenomenon that occurs when one manipulates with the value of $\alpha$. In particular, if $\alpha$ becomes less than (\ref{eq99}) for $p=1$, the first zero disappears. In other words, we loose the first  eigenfunction completely, an eigenfunction  which determine the first term in the leaky mode expansion. Since one usually expects that the first terms in the expansion are the most important ones for generic functions, the loss off the first eigenfunction is ominous. We conjecture that this loss, at least partly, explains why the leaky mode expansion loses its ability to accurately represent important boundary data like a Gaussian,  when the parameter $\alpha$ become small enough. 

\begin{figure}
  \centering
\captionsetup{width=0.85\textwidth}
  \includegraphics[scale=0.6]{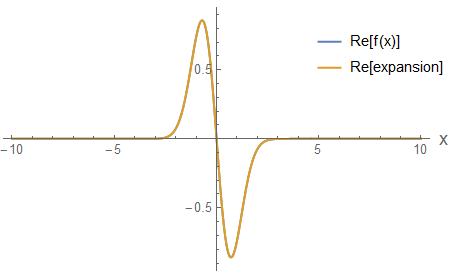}
  \caption{The test function in this figure is $f(x)=\text{exp}(-x^2)(-2x)$. Parameters used in this expansion were $a=10,\omega/c\approx 1.58153\times 10^7,n=1+10^{-24}$. }
\label{fig18}
\end{figure}

In support of this conjecture, note  that the eigenfunctions $\psi^-_p(x)$ alternate between being odd and even functions depending on the index $p$. Before the disappearing of the first zero, $\psi_1^-(x)$ is an even function. Let us denote (\ref{eq99}) for $p=1$ as $\alpha^*$. Then for $\alpha\gtrsim\alpha^*$ the expansion is a good representation of the Gaussian, which is even. However, for $\alpha<\alpha^*$ the expansion goes bad because we have lost the first term in the sum. Now we understand why the expansion goes wrong. Because the most important first term in the expansion is an odd function trying to represent an even Gaussian. With this in mind, let's expand an odd function instead of the even Gaussian for $\alpha<\alpha^*$. Let us for example use derivative of the Gaussian. And indeed, as can be seen in figure (\ref{fig18}), the leaky mode expansion represent the odd functions much better than the even Gaussian. While the Gaussian was badly represented by its leaky mode expansion  for $n=1+10^{-18}$,  for the derivative of the Gaussian, which is an odd function,  we have a very precise leaky mode expansion,  even for an index step as small as $n=1+10^{-24}$.

\section{Asymptotic series}
In the previous section we conjectured that for small  index steps, the leaky mode expansions does converge, but not to the functions used to generate the expansion. We used high precision numerical calculations to support this conjecture.  In this section we will lend  additional support to the conjecture by proving that in the limit of small index step, the leaky mode expansion does indeed converge point wise, but to the wrong function. The asymptotic regime we are exploring are more here conveniently defined in terms of the parameter $\alpha$. The requirement of the analysis in this section is that 
\begin{align}
\sqrt{\alpha}\ll |\xi_{0p}|.\label{AsymptoticCondition}
\end{align}
The validity of this inequality is what we in this section  mean by the asymptotic limit. Note that (\ref{AsymptoticCondition}) is in fact also the requirement for the formula (\ref{eq70}) to be an accurate approximation to the locating of the zeros of the determinant $\xi_{0p}$.
  Recall that the leaky mode expansion for some function $f(x)$ is given by 
\begin{align}
f(x)&=\sum_{p=1}^\infty\frac{\left(f(x),\psi^-_{\xi_{0p}}(x)\right)^-}{\left(\psi^-_{\xi_{0p}}(x),\psi^-_{\xi_{0p}}(x)\right)^-}\psi^-_{\xi_{0p}}(x).\label{eq70}
\end{align}
 We will be interested in finding an  asymptotic approximation to the terms in this sum for two sample functions. However, before we proceed to the actual terms for our sample functions, we first need to know, how the coefficients $A,B,C,D$ in eq. (\ref{eq50}) depend on  the index $p$ in the asymptotic limit. First of all, we realize that the vector $(A,B,C,D)^T$ is the null space and thus eigenvector belonging to the eigenvalue 0. To compute this eigenvector we can proceed, as we would normally do,  by row-reducing the matrix equation (\ref{eq18}), which gives us the matrix 
\begin{align}
\begin{pmatrix}
 e^{i a \xi } & -e^{-i a \xi_0} & -e^{i a \xi_0} & 0 \\
 0 & -i e^{-i a \xi_0} (\xi +\xi_0) & i e^{i a \xi_0} (\xi_0-\xi ) & 0 \\
 0 & 0 & i e^{-3 i a \xi_0} \left(\xi +\xi_0+e^{4 i a \xi_0} (\xi_0-\xi )\right) & -i e^{i a (\xi -2 \xi_0)} (\xi +\xi_0) \\
 0 & 0 & 0 & \eta(\xi_0) \\
\end{pmatrix},\label{eq71}
\end{align}
where
\begin{align}
\eta(\xi_0)=-\frac{i (\xi +\xi_0) e^{i a (\xi -2 \xi_0)} \left(\xi ^2 \left(-1+e^{4 i a \xi_0}\right)-2 \xi  \xi_0 \left(1+e^{4 i a \xi_0}\right)+\xi_0^2 \left(-1+e^{4 i a \xi_0}\right)\right)}{\xi_0 \left(\xi  e^{4 i a \xi_0}-\xi_0 e^{4 i a \xi_0}+\xi +\xi_0\right)}.\label{eq72}
\end{align}
Observe that $\eta(\xi_0)$ contains the determinant of $\textbf{M}$ in the numerator, so we get all zeros in the last row of the matrix,  if $\xi_0=\xi_{0p}$. Using this simplification it is easy to find a basis for the one dimensional null space in the form
\begin{align}
\begin{pmatrix}
A\\
B\\
C\\
D
\end{pmatrix}=
\begin{pmatrix}
\frac{2 \xi_0 e^{2 i a \xi_0}}{\xi  \left(1-e^{4 i a \xi_0}\right)+\xi_0\left(1+ e^{4 i a \xi_0}\right)}\\
-\frac{(\xi -\xi_0) e^{i a (\xi +3 \xi_0)}}{\xi  \left(1-e^{4 i a \xi_0}\right)+\xi_0\left(1+ e^{4 i a \xi_0}\right)}\\
\frac{(\xi +\xi_0) e^{i a (\xi +\xi_0)}}{\xi  \left(1-e^{4 i a \xi_0}\right)+\xi_0\left(1+ e^{4 i a \xi_0}\right)}\\
1
\end{pmatrix}.\label{eq73}
\end{align}
Using the asymptotic expression (\ref{eq28}), for the location of the zeros of the determinant $\xi_0=\xi_{0p}$, we  find the asymptotic formula for the term $\text{exp}(\pm ia\xi_{0p})$ in the following form
\begin{align}
\text{exp}(\pm ia\xi_{0p})&=\text{exp}\left[\pm ia\left(-\frac{p\pi}{2a}+i\frac{1}{4a}\text{Log}\left(\frac{\pi^4p^4}{\alpha^2a^4}\right)\right)\right]=\text{exp}\left[\mp i\frac{p\pi}{2}\right]\left(\frac{\pi p}{\sqrt{\alpha} a}\right)^{\mp 1}.\label{eq74}
\end{align}
So the term $\text{exp}(-ia\xi_{0p})$ grows linearly in $p$, while $\text{exp}(+ia\xi_{0p})$ decays. This helps us write approximate (\ref{eq73}) by the expression
\begin{align}
\begin{pmatrix}
A\\
B\\
C\\
D
\end{pmatrix}\approx
\begin{pmatrix}
\frac{2 \xi_0 e^{2 i a \xi_0}}{\xi+\xi_0}\\
-\frac{(\xi -\xi_0) e^{i a (\xi +3 \xi_0)}}{\xi+\xi_0}\\
\frac{(\xi +\xi_0) e^{i a (\xi +\xi_0)}}{\xi+\xi_0}\\
1
\end{pmatrix}=\begin{pmatrix}
\frac{2 \xi_0 e^{2 i a \xi_0}}{\xi+\xi_0}\\
-\frac{(\xi -\xi_0) e^{i a (\xi +3 \xi_0)}}{\xi+\xi_0}\\
e^{i a (\xi +\xi_0)}\\
1
\end{pmatrix}.\label{eq75}
\end{align}
In the asymptotic limit  we also clearly have 
\begin{align}
\xi+\xi_0&=\sqrt{\xi_0^2+\alpha}+\xi_0=-\xi_0\sqrt{1+\frac{\alpha}{\xi_0^2}}+\xi_0\approx-\xi_0\left(1+\frac{\alpha}{2\xi_0^2}\right)+\xi_0=-\frac{\alpha}{2\xi_0^2},\label{eq76}\\
\xi-\xi_0&=\sqrt{\xi_0^2+\alpha}-\xi_0=-\xi_0\sqrt{1+\frac{\alpha}{\xi_0^2}}-\xi_0\approx-\xi_0\left(1+\frac{\alpha}{2\xi_0^2}\right)-\xi_0\approx-2\xi_0,\label{eq77}\\
\xi+3\xi_0&=\sqrt{\xi_0^2+\alpha}+3\xi_0=-\xi_0\sqrt{1+\frac{\alpha}{\xi_0^2}}+3\xi_0\approx-\xi_0\left(1+\frac{\alpha}{2\xi_0^2}\right)+3\xi_0\approx2\xi_0,\label{eq78}
\end{align}
so that equation (\ref{eq75}) can be further simplified into
\begin{align}
\begin{pmatrix}
A\\
B\\
C\\
D
\end{pmatrix}\approx
\begin{pmatrix}
-\frac{4 \xi_0^2 e^{2 i a \xi_0}}{\alpha }\\
-\frac{4 \xi_0^2 e^{2 i a \xi_0}}{\alpha }\\
e^{-\frac{i a \alpha }{2 \xi_0}}\\
1
\end{pmatrix}\approx\begin{pmatrix}
-\frac{4 \xi_0^2 e^{2 i a \xi_0}}{\alpha }\\
-\frac{4 \xi_0^2 e^{2 i a \xi_0}}{\alpha }\\
1\\
1
\end{pmatrix}\approx
\begin{pmatrix}
(-1)^{p+1}\\
(-1)^{p+1}\\
1\\
1
\end{pmatrix}.\label{eq79}
\end{align}

For any given function $f(x)$ we can write the leaky mode expansion in the form 
\begin{align}
f(x)&=\sum_p\frac{b_p(x)}{N_p}=\sum_pc_p(x),\label{eq80}
\end{align}
where according to eq. (\ref{eq58})
\begin{align}
b_p(x)&=\int_{-a}^af(x)\psi_{p}^-(x)\mathrm{d}x\psi_{p}^-(x),\label{eq81}\\
N_p&=\left( i\int_{-\infty}^{-a}+\int_{-a}^{a}+i\int_{a}^{\infty}\right)\left(\psi_{p}^-(x)\right)^2\mathrm{d}x.\label{eq82}
\end{align}
We now turn our attention to the normalization terms $N_p$ in equation (\ref{eq82}). In the limit of small $\alpha$  since the zeros $\xi_{0p}$ in eq. (\ref{eq70}) are from the 2$^{st}$ quadrant, we can assume $\xi_p\approx-\xi_{0p}$. Using this assumption, and doing the integrals in (\ref{eq82}) exactly, we obtain
\begin{align}
N_p&\approx\frac{k}{\xi_{0p}}\left(\frac{\text{exp}(-2ia\xi_{0p})}{2i}+\frac{4aBC\xi_{0p}}{k}+\sin(2a\xi_{0p})\right)\nonumber\\
&=\frac{k}{\xi_{0p}}\left(\frac{4aBC\xi_{0p}}{k}+\frac{\text{exp}(-2ia\xi_{0p})}{2i}+\frac{\text{exp}(2ia\xi_{0p})}{2i}-\frac{\text{exp}(-2ia\xi_{0p})}{2i}\right)\nonumber\\
&=\frac{k}{\xi_{0p}}\left(\frac{4aBC\xi_{0p}}{k}+\frac{\text{exp}(2ia\xi_{0p})}{2i}\right),\label{eq83}
\end{align}
where $k=A^2+D^2=B^2+C^2$. We know from eq. (\ref{eq74}) that $\text{exp}(2ia\xi_{0p})$ decays as $p^{-2}$, thus the whole expression (\ref{eq83}) simplifies into
\begin{align}
N_p\approx4aBC\approx4a(-1)^{p+1}.\label{eq84}
\end{align}

\noindent  Let us next make a general statement about the decay rate, as a function of the index $p$, of the projection (\ref{eq81}) of a given function $f(x)$ onto the leaky mode $\psi^-_p(x)$. Let us assume that $f(x)$  is  a function that is zero at $x=\pm a$ and is $n$-times continuously differentiable. It is clear that each time we perform integration by parts in  (\ref{eq81}), i.e. differentiating $f(x)$ and integrating $\psi^-_p(x)$, we get an extra factor $i\xi_{0p}$ in the denominator. After $n$ consecutive integrations by parts we get
\begin{align}
b_p(x)=\psi^-_p(x)\left(\frac{i}{\xi_{0p}}\right)^n\int_{-d}^d f^{(n)}(x)\left(\psi^-_p(x)\right)^{\bigstar(n)}\mathrm{d}x,\label{eq85}
\end{align}
where
\begin{align}
\left(\psi^-_p(x)\right)^{\bigstar(n)}&=\left\{
\begin{array}{cc}
B\exp(i\xi_{0p}x)+C\exp(-i\xi_{0p}x) & n\text{ is even}\\
B\exp(i\xi_{0p}x)-C\exp(-i\xi_{0p}x) & n\text{ is odd}
\end{array}\right..\label{eq86}
\end{align}
The term $1/\xi_{0p}^n$ can be approximated as $\approx(-2a)^n/(p\pi)^n$, so the asymptotic expression for eq. (\ref{eq85}) becomes
\begin{align}
b_p(x)&\approx  \psi^-_p(x)\left(\frac{-2ai}{p\pi}\right)^n\int_{-d}^d f^{(n)}(x)\left(\psi^-_p(x)\right)^{\bigstar(n)}\mathrm{d}x.\label{eq87}
\end{align}

\begin{figure}[t]
  \centering
\captionsetup{width=0.85\textwidth}
\begin{subfigure}{.5\textwidth}
  \centering
  \includegraphics[scale=0.35]{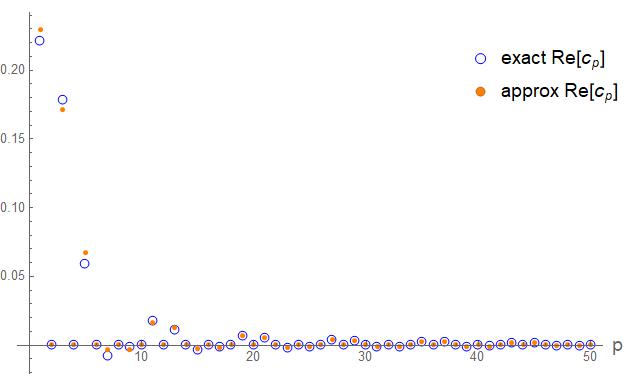}
  \caption{$x=0$}
  \label{fig13a}
\end{subfigure}%
\begin{subfigure}{.5\textwidth}
  \centering
  \includegraphics[scale=0.35]{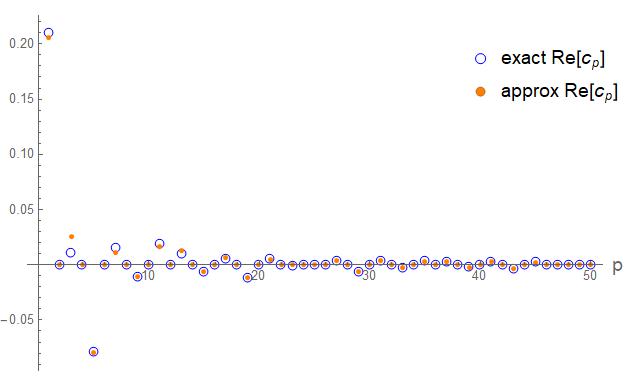}
  \caption{$x=1/3$}
  \label{fig13b}
\end{subfigure}
\caption{Comparing the real part of the exact values of $c_p(x)$ with their asymptotic forms. The parameters used in these plots were $a=1,\omega/c\approx 1.58153\times 10^7,n=1+10^{-14},\alpha\approx 5,d=1/2$. The test function for these coefficients is eq. (\ref{eq88}).}
\label{fig13}
\end{figure}
\noindent
Next, we will find asymptotic approximations to (\ref{eq87}) for  our two chosen sample functions. This will give us an asymptotic approximation to the terms of the leaky mode expansion for the two sample functions.

\begin{figure}[h!]
  \centering
\captionsetup{width=0.85\textwidth}
  \includegraphics[scale=0.6]{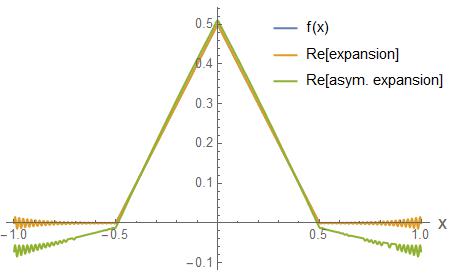}
  \caption{Comparing the original test function eq. (\ref{eq88}) with its leaky modes expansion and the asymptotic leaky modes expansion. Parameters used in this expansion were $a=1,\omega/c\approx 1.58153\times 10^7,n=1+10^{-14},\alpha\approx 5,d=1/2$.}
\label{fig14}
\end{figure}

As our first sample we choose the following triangle function
\begin{align}
f(x)=\left\{
\begin{array}{cc}
x+d & -d<x<0\\
-x+d & 0<x<d
\end{array}\right.,\label{eq88}
\end{align}
whose derivative is 1 for $-d<x<0$ and -1 for $0<x<d$. In this case, the approximative coefficients $c_p(x)=b_p(x)/N_p$ after one integration by parts are
\begin{align}
c_p(x)&=\frac{1}{4a(-1)^{p+1}}\psi^-_p(x)\left(\frac{-2ai}{p\pi}\right)\left(\int_{-d}^0 1\cdot\left(\psi^-_p(x)\right)^{\bigstar(1)}\mathrm{d}x+\int_{0}^d (-1)\cdot\left(\psi^-_p(x)\right)^{\bigstar(1)}\mathrm{d}x\right)\nonumber\\
&=\frac{1}{4a(-1)^{p+1}}\psi^-_p(x)\left(\frac{-2ai}{p\pi}\right)\left(\int_{-d}^0 B\exp(i\xi_{0p}x)-C\exp(-i\xi_{0p}x)\mathrm{d}x\right.\nonumber\\
&\left.-\int_{0}^d B\exp(i\xi_{0p}x)-C\exp(-i\xi_{0p}x)\mathrm{d}x\right)\nonumber\\
&=\frac{1}{4a(-1)^{p+1}}\psi^-_p(x)\left(\frac{-2ai}{p\pi}\right)\frac{i(B+C)}{\xi_0}(\text{exp}(id\xi_0)+\text{exp}(-id\xi_0)-2)\nonumber\\
&=\frac{(-1)^{p+1}+1}{4a(-1)^{p+1}}\psi^-_p(x)\left(\frac{-2ai}{p\pi}\right)^2(\text{exp}(id\xi_0)+\text{exp}(-id\xi_0)-2).\label{eq89}
\end{align}
Using the approximations from (\ref{eq74}),  we get
\begin{align}
c_p(x)&\approx-\frac{4 a^2}{(p\pi)^2} \frac{\left((-1)^p-1\right)}{4a(-1)^{p+1}} \left(-1+ \text{exp}\left(\frac{i \pi  d p}{2 a}\right) \left(\frac{p\pi}{a \sqrt{\alpha }}\right)^{d/a}\right)^2 \nonumber\\
&\text{exp}\left(-\frac{i \pi  p (d+x)}{2 a}\right) \left((-1)^p- \text{exp}\left(\frac{i \pi  p x}{a}\right) \left(\frac{p\pi}{a \sqrt{\alpha }}\right)^{\frac{2 x}{a}}\right) \left(\frac{p\pi}{a \sqrt{\alpha }}\right)^{-\frac{d+x}{a}}.\label{eq90}
\end{align}
In figures (\ref{fig13}) and (\ref{fig14}) we compare the asymptotic expressions for the terms in the leaky mode expansion with the exact terms calculated using high precision numerics. As we can see, there is a remarkable agreement between the values predicted by the asymptotic formulas and the exact values, even for small values of the mode  index $p$. Numerically, the terms appear to approach zero fairly quickly, indicating the the series itself converge. 

In order to see if, and for which values of $x$ the series converge or diverge, we write the terms in the series (\ref{eq90}) into the following form
\begin{align}
c_p(x)&=2 a (-1)^{-p} \left((-1)^p-1\right) e^{\frac{i \pi  p x}{2 a}} \left(a \sqrt{\alpha }\right)^{-x/a}(p\pi)^{x/a-2}-2 a \left((-1)^p-1\right) e^{-\frac{i \pi  p x}{2 a}}\left(a\sqrt{\alpha}\right)^{x/a}(p\pi)^{-x/a-2}\nonumber\\
&+a \left((-1)^p-1\right) e^{-\frac{i \pi  p (d+x)}{2 a}} \left(a \sqrt{\alpha }\right)^{\frac{d+x}{a}}(p\pi)^{-\frac{d+x}{a}-2}+a \left((-1)^p-1\right) e^{\frac{i \pi  p (d-x)}{2 a}} \left(a \sqrt{\alpha }\right)^{-\frac{d-x}{a}}(p\pi)^{\frac{d-x}{a}-2}\nonumber\\
&+a (-1)^{1-p} \left((-1)^p-1\right) e^{\frac{i \pi  p (x-d)}{2 a}} \left(a \sqrt{\alpha }\right)^{-\frac{x-d}{a}}(p\pi)^{\frac{x-d}{a}-2}\nonumber\\
&+a (-1)^{1-p} \left((-1)^p-1\right) e^{\frac{i \pi  p (d+x)}{2 a}} \left(a \sqrt{\alpha }\right)^{-\frac{d+x}{a}}(p\pi)^{\frac{d+x}{a}-2}.\label{eq91}
\end{align}

\noindent

The sum over $p$ of each term in (\ref{eq91}) can be expressed  using the polylogarithm function $Li(n,z)$, which is defined by the expression
\begin{align}
Li(n,z)&=\sum_{p=1}^\infty\frac{z^p}{p^n}.\label{eq92}
\end{align}
Looking at the form of the exponents of the terms in $c_p(x)$, the values of $n$ occurring in the polylogarithms needed to sum all the terms, are $2\pm x/a,2\pm (d+x)/a$ and $2\pm (d-x)a$. If $n$ is strictly larger than 1, the series defining the polylogarithm converge absolutely.  After analyzing the various inequalities we find that we get absolute convergence of the leaky mode expansion for the triangle function only if  $-a+d<x<a-d$. For the case in figure (\ref{fig14}) this region is $-1/2<x<1/2$. However, we also  have a convergence in the entire channel.  In the region outside the region of absolute convergence, we have also have convergence. The convergence here  is ensured by cancellations among terms spiraling towards  the origin in the complex plane.  In the region of the channel outside the domain $-a+d<x<a-d$,  the amplitude of the terms does not decay fast enough ensure absolute convergence and the cancellation among the spiraling terms are needed for convergence. The resulting convergence is evidently only conditional. 

\begin{figure}[h!]
  \centering
\captionsetup{width=0.85\textwidth}
  \includegraphics[scale=0.6]{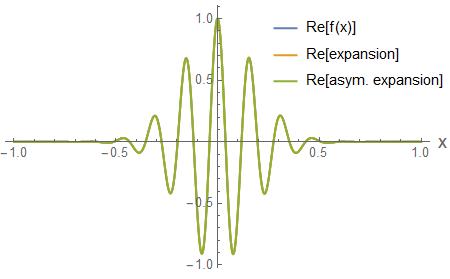}
  \caption{Comparing the original test function eq. (\ref{eq93}) with its leaky modes expansion and the asymptotic leaky modes expansion. Parameters used in this expansion were $a=1,\omega/c\approx 1.58153\times 10^7,n=1+10^{-14},\alpha\approx 5,m=4,k=40$.}
\label{fig15}
\end{figure}
\noindent

As our second sample function we pick a Gaussian wave packet
\begin{align}
f(x)&=\text{exp}(-(mx)^2)\text{exp}(ikx),\label{eq93}
\end{align}
for some real numbers $m,k>0$. The asymptotic terms $c_p(x)$ for this case are 
\begin{align}
c_p(x)&\approx\frac{1}{4a(-1)^{p+1}}\psi^-_p(x)\int_{-a}^a\text{exp}(-(mx)^2)\text{exp}(ikx)\left(B\exp(i\xi_{0p}x)+C\exp(-i\xi_{0p}x)\mathrm{d}x\right)\mathrm{d}x.\label{eq94}
\end{align}
Let the parameters $m,k$ be such that $f(x)$ has its support well inside the slab and $f(\pm a)\approx 0$. Then we can evaluate the integral analytically as
\begin{align}
c_p(x)&\approx\frac{\sqrt{\pi}}{4ma(-1)^{p+1}}\psi^-_p(x)\left(B\text{exp}\left(-\frac{(k+\xi_{0p})^2}{4m^2}\right)+C\text{exp}\left(-\frac{(k-\xi_{0p})^2}{4m^2}\right)\right)\nonumber\\
&\approx\frac{\sqrt{\pi}}{4ma(-1)^{p+1}}\left(B\text{exp}\left(-\frac{(k+\xi_{0p})^2}{4m^2}\right)+C\text{exp}\left(-\frac{(k-\xi_{0p})^2}{4m^2}\right)\right)\nonumber\\
&\left((-1)^{p+1}\text{exp}\left(-i\frac{xp\pi}{2a}\right)\left(\frac{p\pi}{a\sqrt{\alpha}}\right)^{-x/a}+\text{exp}\left(i\frac{xp\pi}{2a}\right)\left(\frac{p\pi}{a\sqrt{\alpha}}\right)^{x/a}\right),\label{eq95}
\end{align}
where $\xi_{0p}$ is defined in (\ref{eq29}).  The terms $c_p(x)$ in this case decay exponentially and thus ensure that the leaky mode expansion converge for all $x$ in the channel. From (\ref{fig15}) we see that the leaky mode expansion and the exact numerical expansion both are very close to the original Gaussian wave packet for all $x$ in the channel.


We have seen that the expansion we introduced in eq. (\ref{eq70}) can represent a function very well as long as the parameter values are not exceeded outside their bounds. These bounds are sufficient for all practical purposes. However, looking at the expansion under such circumstances where the value of $\alpha$ is small enough, we see that the expansion is a very bad representation of the target function. Although we can't state that we  know the reason for this, we have done some preliminary investigations that points to a likely explanation.

 Recall that we do have completeness for the scattering modes. Formally this is expressed by the identity
\begin{align}
\int_{-\infty}^\infty\varphi_{\xi_0}(x)\varphi^*_{\xi_0}(x')\mathrm{d}\xi_0=\delta(x-x').\label{eq100}
\end{align}
Here, $\varphi_{\xi_0}(x)$, can be any linear combination of scattering modes. The usual way to get from the completeness for scattering modes to the completeness for the leaky modes is to analytically extend the scattering modes into the complex frequency space and then use the Cauchy theorem. This allows us to write it as a discrete sum of residues evaluated at the poles which are $\xi_{0j}$. Thus the scattering states get converted into resonant leaky modes at these points leaving us with a sum similar to the one in  to (\ref{eq70}).

\begin{figure}[t]
  \centering
\captionsetup{width=0.85\textwidth}
  \includegraphics[scale=0.4]{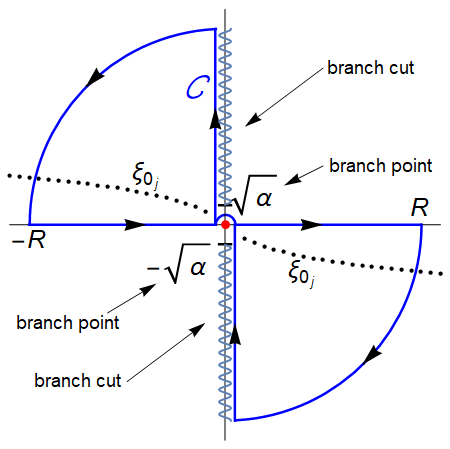}
  \caption{Complex integration contour $\mathcal{C}$.}
\label{fig17}
\end{figure}

In order to be more precise about this, we introduce an integration contour $\mathcal{C}$ in figure (\ref{fig17}) that contains the zeros $\xi_{0j}$ in the second and fourth quadrant. In the scattering states, the continuity coefficients contain the determinant of the matrix (\ref{eq18}), which contains the variable $\xi=\sqrt{\alpha+\xi_0^2}$. This is a complex square root that has a branch cut on the negative real axis. Figure (\ref{fig17}) depicts one possible complex contour. We indicated the branch points, where the branch cut begins. 

Integrating  the integrand in  (\ref{eq100}) over the contour $\mathcal{C}$, we get
\begin{align}
\int_{\mathcal{C}}\varphi_{\xi_0}(x)\varphi^*_{\xi_0}(x')\mathrm{d}\xi_0=\int_{\mathcal{C}_R}+\int_{\mathcal{C}_r}+\int_{\mathcal{C}_i}\varphi_{\xi_0}(x)\varphi^*_{\xi_0}(x')\mathrm{d}\xi_0,\label{eq1006}
\end{align}
where $\mathcal{C}_R$ denotes the circular part of $\mathcal{C}$, $\mathcal{C}_r$ is the contour part along the real axis and $\mathcal{C}_i$ is the one along the imaginary axis. Letting $R\to\infty$ and assuming that the contribution from the integrals over $\mathcal{C}_R$ and $\mathcal{C}_i$ vanish in the limit, we have
\begin{align}
\int_{\mathcal{C}}\varphi_{\xi_0}(x)\varphi^*_{\xi_0}(x')\mathrm{d}\xi_0=\delta(x-x').\label{eq1007}
\end{align}
With the aid of Cauchy theorem we write the left-hand side in (\ref{eq1007}) as
\begin{align}
2\pi i\sum_{j=0}^\infty\text{Res}\left(\varphi_{\xi_0}(x)\varphi^*_{\xi_0}(x'),\xi_{0j}\right)&=\delta(x-x').\label{eq1008}
\end{align}

Let us now obtain the expressions for the scattering states. We can write them in the form
\begin{align}
\psi^{-}_{\xi_0}(x)&=\left\{
\begin{array}{cc}
A^+\text{exp}(i\xi x)+A^-\text{exp}(-i\xi x), & x<-a\\
B\text{exp}(i\xi_0 x)+C\text{exp}(-i\xi_0 x), & -a<x<a\\
D\text{exp}(i\xi x), & a<x
\end{array}\right..\label{eq1009}
\end{align}
With the usual boundary conditions we end up with a system where we have one free parameter. Solving the system leaves us with the following solution
\begin{align}
\begin{pmatrix}
A^-\\
B\\
C\\
D
\end{pmatrix}=\frac{A^+}{\det\textbf{M}}\begin{pmatrix}
 e^{-2 i a \xi_0} \left(-1+e^{4 i a \xi_0}\right) \left(\xi ^2-\xi_0^2\right)\\
-2  \xi  (\xi +\xi_0) e^{i a (\xi -\xi_0)}\\
2  \xi  (\xi -\xi_0) e^{i a (\xi +\xi_0)}\\
-4  \xi  \xi_0
\end{pmatrix},\label{eq1010}
\end{align}
where $\textbf{M}$ is the matrix found in (\ref{eq18}). Once we have obtained the scattering states, we use the relation (\ref{eq1008}) to expand the function $f(x)$ by multiplying both sides with it and integrate wrt $x'$. This gives us the identity
\begin{align}
2\pi i\sum_{j=0}^\infty\lim_{\xi_0\to\xi_{0j}}(\xi_0-\xi_{0j})\varphi_{\xi_0}(x)\int_{-\infty}^\infty f(x')\varphi^*_{\xi_0}(x')\mathrm{d}x'&=f(x).\label{eq1012}
\end{align}
The goal is to match the two sums  (\ref{eq70}) and  (\ref{eq1012}). Comparing them, we observe that the inner product defined as in (\ref{eq58}) does not include any complex conjugate and neither does the sum  (\ref{eq70}). In order to match (\ref{eq70}) and (\ref{eq1012}), we therefore need $\varphi_{\xi_0}(x)$ to be real. One possibility for the integrand in (\ref{eq100}) would be
\begin{align}
\varphi_{\xi_0}(x)\left(\varphi\right)^*_{\xi_0}(x')=\left(\psi^-_{\xi_0}(x)+\overline{\psi^-}_{\xi_0}(x)\right)\overline{\left(\psi^-_{\xi_0}(x')+\overline{\psi^-}_{\xi_0}(x')\right)},\label{eq1013}
\end{align}
where $\overline{\psi^-}_{\xi_0}(x)=\left(\psi^-_{\xi_0^*}(x)\right)^*$ to ensure the analyticity of the state as a function of complex wavenumber $\xi_0$.

The free parameter in  (\ref{eq1010}) controls  which term in the expression  (\ref{eq1013}) are going to have poles at $\xi_{0j}$. In other words, it controls which terms vanish after evaluating the residues. Notice that after evaluating the residues at $\xi_{0j}$, the only survivor we want is $\psi^-_{\xi_0}(x)$. A suitable choice for achieving this is
\begin{align}
A^+&=\overline{\det\textbf{M}}^\frac{1}{2}\det\textbf{M}^\frac{1}{2}a^+(\xi_0).\label{eq1016}
\end{align}
With this choice, the coefficients in  (\ref{eq1010}) have common factor $\overline{\det\textbf{M}}^\frac{1}{2}/\det\textbf{M}^\frac{1}{2}$.  The cross terms from $\varphi_{\xi_0}(x)\left(\varphi\right)^*_{\xi_0}(x')$ vanish leaving us with one term of the form $\overline{\psi^-}_{\xi_0}(x)\overline{\psi^-}_{\xi_0}(x')$ whose factor is $\det\textbf{M}/\overline{\det\textbf{M}}$. Since $\overline{\det\textbf{M}}$ has no zeros at the outgoing $\xi_{0j}$, this terms vanishes too upon taking the residues, leaving only the desired term $\psi^-_{\xi_0}(x)\psi^-_{\xi_0}(x')$.

Thus, with these choices made, the completeness of the scattering states implies that the following identity holds

\begin{align}
2\pi i\sum_{j=0}^\infty\lim_{\xi_0\to\xi_{0j}}(\xi_0-\xi_{0j})\psi^-_{\xi_0}(x)\int_{-\infty}^\infty f(x')\psi^-_{\xi_0}(x')\mathrm{d}x'&=f(x).\label{eq1015}
\end{align}
We now equate the sums (\ref{eq70}) and (\ref{eq1015}). 
\begin{align}
&\sum_{j=0}^\infty\frac{\left(f(x),\psi^-_{\xi_{0j}}(x)\right)^-}{\left(\psi^-_{\xi_{0j}}(x),\psi^-_{\xi_{0j}}(x)\right)^-}\psi^-_{\xi_{0j}}(x)=2\pi i\sum_{j=1}^\infty\lim_{\xi_0\to\xi_{0j}}(\xi_0-\xi_{0j})\psi^-_{\xi_0}(x)\int_{-a}^a f(x')\psi^-_{\xi_0}(x')\mathrm{d}x',\nonumber\\
\nonumber\\
&\Updownarrow\nonumber\\
\nonumber\\
&\sum_{j=0}^\infty\frac{1}{\left(\psi^-_{\xi_{0j}}(x),\psi^-_{\xi_{0j}}(x)\right)^-}\int_{-a}^a f(x')\psi^-_{\xi_0}(x')\mathrm{d}x'\psi^-_{\xi_{0j}}(x)&\nonumber\\
\nonumber\\
&=2\pi i\sum_{j=1}^\infty\lim_{\xi_0\to\xi_{0j}}\frac{(\xi_0-\xi_{0j})}{\det\textbf{M}}\overline{\det\textbf{M}}(\xi_{0j}) (a^+)^2(\xi_{0j})\int_{-\infty}^\infty f(x')\psi^-_{\xi_{0j}}(x')\mathrm{d}x'\psi^-_{\xi_{0j}}(x),\label{eq1018}
\end{align}
where $\psi^-_{\xi_{0j}}(x)$ is the function defined in (\ref{eq1009}) with $\xi_0\to\xi_{0j}$. In this limit, the right hand side goes to zero and the matrix becomes singular and the  coefficients $(A^-,B,C,D)$ approach a basis for the null space of the matrix which leads the solution to be the nullspace of $\textbf{M}$, so (\ref{eq1009}) becomes the resonant states with the coefficients $(A^-,B,C,D)$ whose definitions are in  (\ref{eq1010}) except of the common factor $1/\det\textbf{M}$ and $a^+$ instead of $A^+$. This factor was modified by (\ref{eq1016}) and put into limit in (\ref{eq1018}). We wrote the inner product of $f(x)$ and the resonant state on the left-hand side as an integral from $-a$ to $a$ because of the compact support of $f(x)$ in this area.

\noindent

It is now evident that we can make the expressions on the two sides of (\ref{eq1018}) identical only  if the free parameter $a^+(\xi_0)$ is chosen to be
\begin{align}
a^+(\xi_{0j})&=\sqrt{\frac{\det'\textbf{M}(\xi_{0j})}{2\pi i\left(\psi^-_{\xi_{0j}}(x),\psi^-_{\xi_{0j}}(x)\right)^-\overline{\det\textbf{M}}(\xi_{0j})}}.\label{eq102}
\end{align}
However, this choice for $a^+(\xi_{0j})$ is not an analytic function because both functions $\det'\textbf{M}(\xi_{0j})$ and $\left(\psi^-_{\xi_{0j}}(x),\psi^-_{\xi_{0j}}(x)\right)^-$ are zero inside the integration contour. Each of these two families of countably many zeros give rise to equally many branch cuts. The parametric formulas for these branch cuts are possible to find but while applying the Cauchy theorem we must now  include terms representing integrals around all these additional branch cuts. Thus what we get from the Cauchy theorem is that any function with compact support inside that channel is equal to its leaky mode expansion, plus additional terms that includes integrals along the branch cuts on and off the imaginary axis as described above. What we know is that, unless alpha is smaller than the critical value $\alpha^*$, which we introduced in section four, the function is well represented by the leaky mode expansion alone. This means that the contribution from all the other terms for such values of alpha are negligible. For smaller values of alpha the contributions from the rest of the terms are not negligible and as $\eta=a^2\alpha$ approaches zero, these terms will come to dominate. For such values of $\alpha$ the leaky mode expansion still converge, but it does not converge to the function used to construct the expansion. By deriving asymptotic formulas for the all the terms defined by integrals around branch cuts,  in the limit when $\eta$ approaches zero, one could compare their sizes and identify the dominant ones. If, say, one term dominate, then this term could be added to the leaky mode expansion resulting in an expansion that represents the function to be expanded in a much better way than the leaky mode expansion is able to do on its own. We believe that the asymptotic expressions for the terms could be found, but there might not be a dominant term, and even if there is, extending the leaky mode expansion by adding this term might easily make the expansion too hard to use for practical calculations.

\section{Conclusion}
In this paper we have presented an new approach to minimizing the reflections from  finite computational boundaries for wave equations formulated in the spectral domain. This approach is based on representing the field in the transverse spatial direction using leaky mode expansions supported by an artificial index channel. We have shown that at the linear level, our approach makes it possible propagate the waves much further than what is possible  if a regular Fourier expansion is used. The leaky modes are not reflectionless at the boundary, and eventually the small but finite reflections build up, and the computed solutions starts to deviate from the infinite domain solutions. This reflection can be  minimized by reducing the index step, but at the price of getting a progressively worse representation of the solution to the wave equation. In section four and five we have argued, using both numerical and analytical approaches, that a practically useful trade off can be made between minimizing reflections from the boundary and maximizing the accuracy of the representation of solutions of the wave equations using leaky modes.

We have illustrated our approach using the case of a TE electromagnetic wave in vacuum, but the approach can clearly be generalized to much more general wave propagation problems than this. In the optical context the obvious next step would be to consider waves with cylinder symmetry. An important issue that we have not discussed in this paper is how to compute the transformation from fields to leaky mode amplitudes and back in an accurate, stable and efficient way. 

\section{Acknowledgments}
The authors are thankful for support from the Department of mathematics and statistics at the
Arctic University of Norway, from the Arizona Center for Mathematical Sciences at the University
of Arizona, and for the support from the Air Force Office for Scientific Research under Grant No.
FA9550-19-1-0032

\bibliographystyle{unsrt}
\bibliography{paper}

\end{document}